\documentclass[sigconf,10pt,nonacm]{acmart}

\renewcommand\footnotetextcopyrightpermission[1]{}
\setcopyright{none}

\makeatletter
\def\@shortauthors{}
\def\@shorttitle{}
\makeatother

\AtBeginDocument{%
  \thispagestyle{plain}
  \pagestyle{plain}
}

\usepackage{algorithm}
\usepackage{algpseudocode}
\usepackage{amsmath}
\usepackage{subcaption} 
\usepackage{pifont}

\usepackage{multirow}           
\usepackage{graphicx}
\usepackage{booktabs}
\usepackage{tabularx}
\usepackage{makecell}
\usepackage{array}
\usepackage{xcolor}
\usepackage{tikz}
\usepackage{float}
\usepackage{titlesec}
\titlespacing{\section}{0pt}{4pt}{4pt}
\titlespacing{\subsection}{0pt}{5pt}{3pt}
\usetikzlibrary{shapes.geometric}

\newcommand{\bstep}[1]{
  \tikz[baseline=(char.base)]{
    \node[shape=circle, fill=black, inner sep=1pt] (char)
    {\color{white}\scriptsize\bfseries #1};
  }
}

\title{A Family of Open Time-Series Foundation Models \\
for the Radio Access Network}

\author{Ioannis Panitsas,  Leandros Tassiulas\\
Department of Electrical and Computer Engineering, Yale University}

\begin{document}

\begin{abstract}
\noindent The Radio Access Network (RAN) is evolving into a programmable and disaggregated infrastructure that increasingly relies on AI-native algorithms for optimization and closed-loop control. However, current RAN intelligence is still largely built from task-specific models tailored to individual functions, resulting in model fragmentation, limited knowledge sharing across tasks, poor generalization, and increased system complexity. To address these limitations, we introduce \textbf{TimeRAN}, a unified multi-task learning framework for time-series modeling in the RAN. \texttt{TimeRAN} leverages a lightweight time-series foundation model with few task-specific heads to learn transferable representations that can be efficiently adapted across diverse tasks with limited supervision. To enable large-scale pretraining, we further curate and open-source \textbf{TimeRAN DataPile}\footnote{The \texttt{TimeRAN DataPile} corpus and the pretrained \texttt{TimeRAN} model weights are available at \url{https://github.com/panitsasi/TimeRAN}}, the largest time-series corpus for RAN analytics to date, comprising over \textbf{355K} time series and \textbf{0.56B} measurements across diverse telemetry sources, protocol layers, and deployment scenarios. We evaluate \texttt{TimeRAN} across a comprehensive set of RAN analytics tasks, including \textit{anomaly detection}, \textit{classification}, \textit{forecasting}, and \textit{imputation}, and show that it achieves state-of-the-art performance with minimal or no task-specific fine-tuning. Finally, we integrate \texttt{TimeRAN} into a proof-of-concept 5G testbed and demonstrate that it operates efficiently with limited resource requirements in real-world scenarios. 
\end{abstract}

\maketitle

\section{INTRODUCTION}

\noindent The Radio Access Network (RAN) is undergoing a fundamental architectural transformation toward 6G, evolving from a monolithic and inflexible system into a disaggregated, multi-purpose infrastructure capable of jointly supporting diverse functionalities, including communication, computation, and sensing \cite{survey1}. In this context, AI-RAN has emerged as a promising paradigm that re-architects the RAN as a programmable computing platform built on general-purpose accelerated hardware capable of supporting both cellular and AI workloads simultaneously \cite{ai_ran_alliance}. Leveraging this architecture, a growing body of work has begun designing, integrating, and evaluating AI-native RAN algorithms to enhance network performance and efficiency. Representative examples include  autoencoder-based techniques for efficient channel estimation \cite{csi_compression}, reinforcement learning–based schedulers for radio resource allocation and control \cite{coloran}, and AI-driven solutions for higher-layer RAN functions such as anomaly detection \cite{fedjam_dataset}, and mobility management \cite{panitsas2024handover}.

Despite these promising advances, several important challenges still remain largely unexplored and unresolved. More specifically, current approaches predominantly focus on designing task-specific learning-based modules for highly specialized RAN functions, leading to a proliferation of models that must be deployed, maintained, and orchestrated across the RAN, thereby increasing system complexity and operational overhead \cite{oran}. In addition, these methods lack mechanisms for effectively sharing learned representations across tasks, requiring separate models per use case and resulting in redundant learning and inefficient use of training and computational resources \cite{coloran}. Along the same lines, current deployed models exhibit limited representation capacity, leading to poor generalization, as models trained in specific operational environments often fail to transfer across distributed cell sites with diverse traffic patterns and radio propagation conditions, frequently requiring full retraining. Finally, these challenges are further exacerbated by the limited computational resources available at the network edge.

A promising direction to address these issues is to leverage \emph{foundation model–inspired paradigms}, which have shown strong capability in learning rich contextual representations and enabling generalization across diverse downstream tasks with limited supervision \cite{foundation_models}. Despite their success across domains, their adoption in telecommunications remains relatively limited. The main reasons are threefold: (i) constrained data availability, as RAN data are often proprietary and not widely accessible \cite{oran}, (ii) the inherent complexity of modeling such data, which consist of stochastic, high-dimensional multivariate time series with varying temporal granularities \cite{telecomts_dataset}, and (iii) limited computational resources at cell sites, which further complicates the deployment of large and computationally intensive foundation models.

Inspired by these limitations, we introduce \texttt{TimeRAN}, a unified multi-task learning framework designed to eliminate model fragmentation, enable cross-task generalization, and reduce system complexity for scalable deployment of learning-based RAN intelligence. \texttt{TimeRAN} employs a lightweight transformer-based time-series architecture to learn generalizable representations from RAN telemetry, coupled with a few task-specific heads for efficient adaptation across diverse network applications. To enable representation learning at scale, we curate and \textit{open-source} the largest time-series corpus for RAN analytics to date, \textit{TimeRAN DataPile} \cite{timeran_datapile}, comprising over 355K time series and 0.56B measurements across diverse telemetry sources, RAN protocol layers, and deployment scenarios, and use it for training. We evaluate \texttt{TimeRAN} across a comprehensive set of RAN analytics tasks, including \emph{anomaly detection}, \emph{classification}, \emph{forecasting}, and \emph{imputation}, achieving state-of-the-art performance with minimal supervision. Finally, we integrate \texttt{TimeRAN} into a proof-of-concept 5G testbed, demonstrating low-latency inference with a limited computational footprint. More specifically, this work makes the following key contributions:

\vspace{-0.5em}

\begin{itemize}
\item We introduce \texttt{TimeRAN}, a unified multi-task learning-based architecture to eliminate task-specific model fragmentation and learn generalizable representations across diverse RAN tasks and deployments.

\item We curate and open-source the \textit{TimeRAN DataPile}, the largest time-series corpus for RAN analytics to date, to enable large-scale pretraining.

\item We evaluate \texttt{TimeRAN} across a comprehensive set of RAN tasks, demonstrating state-of-the-art performance with minimal fine-tuning and supervision.

\item We integrate \texttt{TimeRAN} into an over-the-air 5G testbed, demonstrating low-latency inference with limited compute requirements in real-world scenarios.

\item We release the pretrained \texttt{TimeRAN} weights to enable further research and experimentation.
\end{itemize}

\section{BACKGROUND}

\noindent\textbf{RAN Architecture.} The RAN acts as the wireless interface connecting the User Equipment (UE) to the cellular core network, supporting user-plane data transmission and control plane operations. In 5G, the RAN protocol stack is disaggregated into three main units: the \emph{Radio Unit (RU)}, the \emph{Distributed Unit (DU)}, and the \emph{Centralized Unit (CU)}, enabling flexible functional decomposition and deployment. The RU performs radio-frequency processing and lower Physical Layer (PHY) functions at the cell site, while the DU executes latency-critical baseband processing, including upper PHY operations, along with the Medium Access Control (MAC) and Radio Link Control (RLC) layers. Higher protocol layers, such as the Packet Data Convergence Protocol (PDCP) and Radio Resource Control (RRC), are handled by the CU under more relaxed latency constraints, facilitating centralized coordination and control. This decomposition corresponds to the widely adopted \emph{functional split 7.2} \cite{ngran}. Finally, these units can be virtualized and deployed on commodity infrastructure, enabling flexible placement and improved compute utilization through resource pooling \cite{slicepilot}, while supporting dynamic scaling and efficient resource allocation.

\noindent\textbf{Foundation Models for the RAN.} Foundation models have emerged as a powerful paradigm for learning representations from large-scale data via self-supervised objectives, enabling a wide range of downstream tasks within a unified framework \cite{foundation_models}. Initially developed for natural language processing, this paradigm has since been extended to multiple modalities, including vision, speech, and time-series data \cite{foundation_models_survey}. Regardless of the input modality, foundation models typically follow a common processing pipeline: raw inputs are first converted into modality-specific tokens, then projected into high-dimensional embeddings, and finally processed by stacked transformer layers \cite{vaswani2017attention} comprising self-attention mechanisms and feedforward networks, enabling the modeling of complex contextual dependencies. In RAN systems, where large volumes of heterogeneous, high-dimensional telemetry are continuously generated, such models provide a natural approach for learning generalizable representations.

\section{RELATED WORK}
\noindent RAN intelligence, primarily driven by time-series telemetry, has emerged as an active research area focused on extracting actionable insights from high-dimensional RAN data to enable efficient network monitoring, optimization, and control \cite{learning_algorithms_survey}. Common approaches include statistical \cite{statistical}, reconstruction-based \cite{spotlight_dataset}, and sequential deep learning models \cite{forecasting1, jamshield_dataset, forecasting_2, panitsas2024handover, hetnets_dataset, blt_dataset} for anomaly detection and forecasting; deep neural network architectures for classification \cite{fedjam_dataset, jamshield_dataset, coloran, tractor_dataset, open_ireland_dataset}; and reconstruction  methods for imputation \cite{statistical, spotlight_dataset, imputation1}. Despite their competitive performance, these models are often tightly coupled to specific operational environments, requiring continuous retraining or fine-tuning under changing conditions, resulting in substantial engineering overhead and maintenance costs \cite{coloran, oran}. Recent work has begun exploring unified learning paradigms, including foundation-model-based approaches, to learn transferable representations \cite{llm_1, llm_2, telecomts_dataset}. Preliminary results demonstrate promising performance across a relatively narrow set of tasks. However, existing works are not specifically designed for RAN settings and remain limited in scope, failing to systematically examine whether and how such models can be effectively adapted to the unique characteristics of RAN. Therefore, this gap motivates the need for unified  models that support scalable and generalizable learning in the RAN.

\section{TIMERAN DATAPILE}

\noindent To address the lack of large-scale data for pretraining and representation learning in the RAN domain, we collate multiple telemetry datasets from widely used public repositories and augment them to form a unified and comprehensive corpus, which we refer to as the \textit{TimeRAN DataPile}\footnote{We envision \textit{TimeRAN DataPile} as a reference corpus for RAN downstream tasks, analogous in spirit to foundational datasets such as ImageNet \cite{imagenet} in computer vision and The Pile \cite{thepile} in natural language processing.}. Our corpus supports a broad set of learning tasks, including \textit{anomaly detection, classification, imputation, and forecasting}, while providing time series measurements spanning the entire RAN and UE protocol stacks. It comprises approximately \textit{29\,GB} of telemetry data, including \textit{355K} unique time series and \textit{0.56\,billion} timestamps collected from both operational cellular networks and over-the-air experimental testbeds.

The \textit{TimeRAN DataPile} exhibits diversity in both channel\footnote{Throughout this paper, we use the term \textit{channel} to denote an individual variable (i.e., one variate) of a multivariate time series.} dimensionality and temporal granularity. Channel dimensionality ranges from a few variables to high-dimensional telemetry with hundreds of variables, reflecting the heterogeneity across RAN deployments and  configurations. Temporal resolution spans multiple monitoring scales, from fine-grained millisecond-level measurements to coarser second- and minute-level intervals, capturing both short-term dynamics and long-term trends. Time-series lengths also vary significantly, ranging from short traces with limited observations to long sequences comprising millions of samples. In terms of dataset composition, \textit{TimeRAN DataPile} includes 9 anomaly detection datasets comprising 23M observations across 21K unique time series; 7 classification datasets with 5M observations across 248 time series, covering tasks such as mobility classification, service identification, congestion detection, traffic classification, and root-cause analysis; and 25 datasets for imputation and forecasting, forming the largest portion of the corpus, with 269M observations across 167K unique time series. Further analytical details on the corpus, curation, and processing pipeline are provided in \cite{timeran_datapile}.

\begin{figure}[t] 
    \centering 
    \includegraphics[width=\columnwidth]{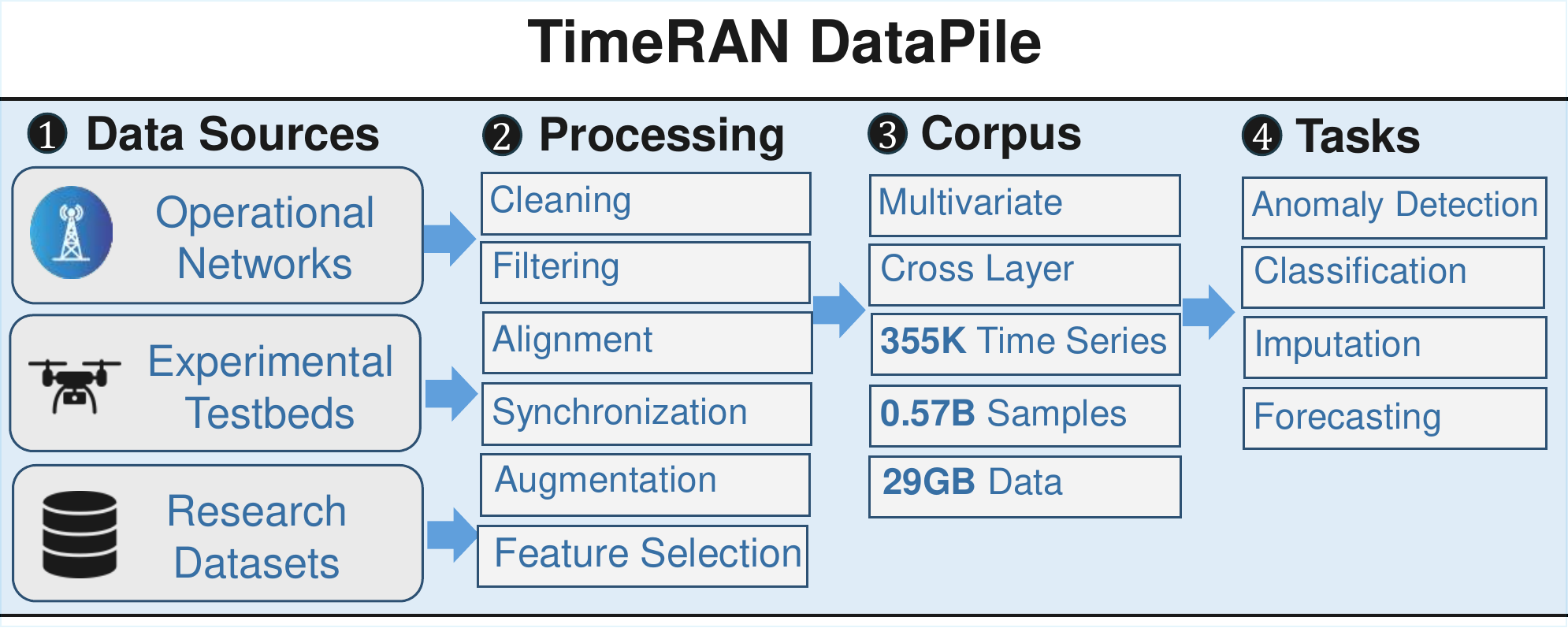} 
    \vspace{-2em}
    \caption{Overview of the TimeRAN DataPile.} 
    \label{fig:timeran_datapile} 
    \vspace{-2em}
\end{figure}

\subsection{Curation and Preprocessing Pipeline}

\noindent Constructing the \texttt{TimeRAN DataPile} is non-trivial due to the heterogeneity of publicly available RAN telemetry datasets. The collected sources differ substantially in protocol-layer coverage, channel dimensionality, temporal granularity, and task coverage. To address these challenges and enable reliable downstream learning, we design a systematic curation and preprocessing pipeline consisting of the following steps:

\noindent\textbf{Channel filtering.}
We retain only numeric measurements directly reflecting RAN behavior and remove auxiliary metadata, categorical descriptors, and identifiers that are not informative for temporal modeling.

\noindent\textbf{Temporal consistency.}
Several sources distribute telemetry across multiple files or partitions. In such cases, we retain only segments that can be reliably time-synchronized. Measurements with inconsistent sampling rates or missing timestamps are discarded to avoid temporal misalignment.

\noindent\textbf{Channel pruning.}
We remove channels that are constant, near-constant, or highly redundant through correlation analysis, as such channels increase dimensionality without contributing meaningful learning signals.

\noindent\textbf{Handling sparse traces.}
For datasets with a limited number of samples (less than 1000), we apply interpolation to increase temporal continuity while preserving the underlying temporal dynamics.

\noindent\textbf{Synthetic anomaly augmentation.}
For a subset of anomaly detection datasets with limited or no anomalous events, we introduce short-duration synthetic anomalies to create or augment anomalous segments while preserving the statistical characteristics of the traces. These anomalies emulate transient network disruptions (e.g., spikes, sudden drops, level shifts, variance changes, and temporary saturation) affecting individual indicators or broader telemetry segments.

\section{TIMERAN DESIGN \& ARCHITECTURE}
\vspace{-0.3em}
\subsection{Overview}

\noindent Building on the large-scale \textit{TimeRAN DataPile}, we next introduce \texttt{TimeRAN}, a unified learning-based framework that supports a broad range of RAN tasks through a shared, lightweight multi-task architecture, focusing on higher-layer RAN functions that do not require strict real-time inference.

\begin{figure}[t]
    \centering
    \includegraphics[width=\columnwidth]{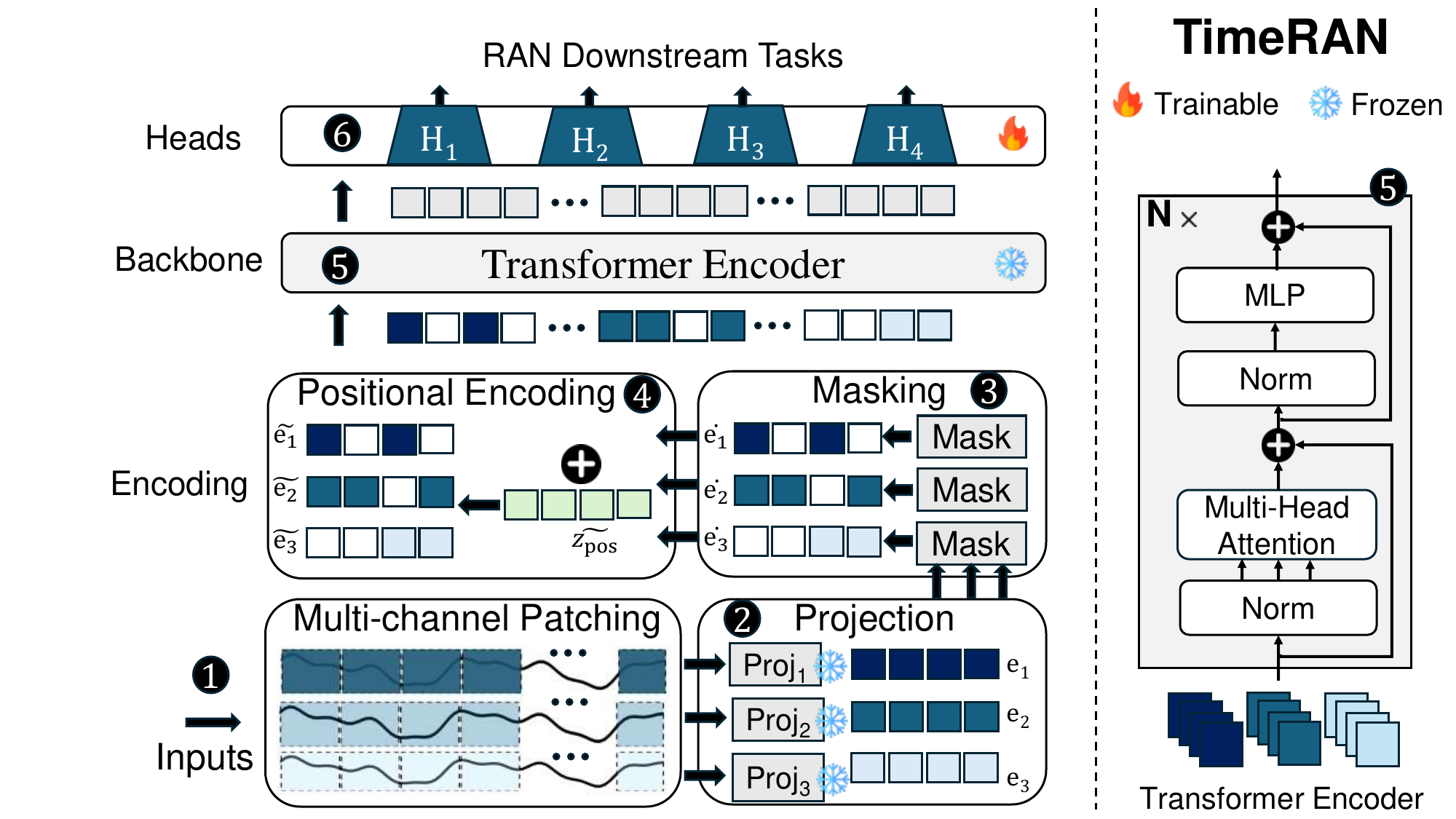}
    \vspace{-2em}
    \caption{TimeRAN system architecture.}
    \label{fig:architecture}
    \vspace{-2.2em}
\end{figure}

As illustrated in Fig.~\ref{fig:architecture}, \texttt{TimeRAN} consists of multiple modules including:
\bstep{1} a \emph{Multi-channel Patching} module that segments multivariate RAN telemetry into fixed-length temporal patches;
\bstep{2} a \emph{Projection} module that maps the patches into a shared latent representation space;
\bstep{3} a \emph{Masking} module, activated only during pre-training, that enables self-supervised learning by selectively masking portions of the patches;
\bstep{4} a \emph{Positional Encoding} module that preserves the temporal ordering of the patches;
\bstep{5} a \emph{Transformer Encoder} backbone that captures temporal dependencies and cross-channel correlations through stacked self-attention layers; and
\bstep{6} a set of \emph{Task-Specific Heads} that operate on the shared latent representation to enable diverse RAN analytics tasks.

\subsection{Design Requirements}
\noindent We design \textit{TimeRAN} with the following key requirements.

\noindent \textbf{1) Unified multi-task learning.}  
The system should follow a \emph{“one encoder, many tasks”} paradigm, where a shared backbone is reused across many downstream tasks.

\noindent \textbf{2) Generalization and efficient adaptation.}  
The framework should learn robust representations that generalize across heterogeneous and unseen environments, while enabling efficient adaptation to new tasks through zero-shot or lightweight fine-tuning without retraining from scratch.

\noindent \textbf{3) Computational efficiency.}  
The system should operate efficiently in operational deployments, such as at the network edge or at cell sites with limited computational resources.

\subsection{TimeRAN Architecture}
\label{sec:timeran_architecture}

\noindent \texttt{TimeRAN} adapts a modular architecture. More specifically:

\noindent{\textbf{5.3.1\quad Patching and Projection Modules.}}
\label{sec:patching}
The first components of \texttt{TimeRAN} are the \textit{Patching and Projection} modules, which convert multivariate RAN telemetry into high-dimensional embeddings that serve as input tokens for the subsequent modules. More specifically, let $\mathbf{X} \in \mathbb{R}^{C \times T}$ denote the telemetry collected over an observation window of length $T$, where $C$ represents the number of channels. First, each channel is partitioned into non-overlapping temporal windows of length $P$, referred to as \textit{patches}. Accordingly, each channel $c \in \{1,\ldots,C\}$ is divided into $N=\lfloor T/P \rfloor$ patches, where the $i$-th patch is defined as $\mathbf{x}_c^{(i)} \in \mathbb{R}^{1 \times P}$.  Next, each patch is normalized and projected into a $d$-dimensional latent space through a trainable linear projection layer, producing an embedding $\mathbf{e}_c^{(i)} \in \mathbb{R}^{1 \times d}$.  Finally, the resulting embeddings across all channels are organized into a sequence $\mathbf{E} \in \mathbb{R}^{(C \cdot N) \times d}$, which is then forwarded to the next module.

\noindent{\textbf{5.3.2\quad Positional Encoding Module.}}
To preserve temporal ordering, positional embeddings are added to the patch embeddings before being processed by the \textit{Transformer Encoder}. Specifically, this module takes as input a sequence of embeddings $\mathbf{E} \in \mathbb{R}^{(C \cdot N) \times d}$ and adds a positional embedding to each patch embedding to inject temporal order information. The resulting position-aware embeddings are then forwarded to the \textit{Time-Series Transformer Encoder}.

\noindent{\textbf{5.3.3\quad Time-Series Transformer Encoder.}}
\label{sec:encoder}
Building on the previous stages, the \textit{Time-Series Transformer Encoder} takes as input the position-aware embeddings and processes them to model temporal dependencies and cross-channel correlations in the multivariate input telemetry. \texttt{TimeRAN} employs a lightweight encoder based on the vanilla Transformer \cite{vaswani2017attention}, adapted for time-series data. More precisely, the encoder consists of $L$ stacked transformer layers, each comprising multi-head self-attention followed by a position-wise feed-forward network, with residual connections and normalization. Within each layer, the input representation $\mathbf{E}$ is projected into queries, keys, and values as $Q = \mathbf{E}W_Q$, $K = \mathbf{E}W_K$, and $V = \mathbf{E}W_V$. Scaled dot-product attention is then computed as $\mathrm{Attn}(Q,K,V)=\mathrm{softmax}\!\left(\frac{QK^\top}{\sqrt{d}}\right)V$. To capture diverse temporal dependencies, the model employs $B$ attention heads operating in parallel on different representation subspaces; the outputs of these heads are concatenated and linearly projected to form the final attention representation. Through this attention mechanism, each embedding attends to all others in the sequence, enabling the model to capture long-term temporal dependencies across the observation window. The encoder therefore produces contextualized representations for all embeddings, denoted as $\mathbf{Z} \in \mathbb{R}^{(C \cdot N) \times d}$.

\noindent{\textbf{5.3.4\quad Task-Specific Heads.}}
\label{sec:heads}
Finally, the contextualized patch embeddings produced by the \textit{Time-Series Transformer Encoder} are consumed by lightweight task-specific heads to enable multiple downstream tasks. \texttt{TimeRAN} supports four time-series tasks, namely \textit{anomaly detection}, \textit{classification}, \textit{forecasting}, and \textit{imputation}, and employs specialized heads for each objective.  All heads share a similar architecture, realized as a stack of linear layers $M$ that operate on the \textit{Transformer Encoder} outputs to produce task-specific predictions. For classification tasks, a global representation is first obtained by aggregating the contextualized patch embeddings. Since no dedicated summary token (e.g., a [CLS] token \cite{bert}) is used, the global representation is computed via mean aggregation as $z = \frac{1}{C \cdot N}\sum_{i=1}^{C \cdot N} Z_i \in \mathbb{R}^{1 \times d}$, which is then used by the classification head to learn a mapping to the target classes.  In contrast, the remaining heads operate directly on the full sequence of contextualized patch embeddings.

\noindent{\textbf{5.3.5\quad Self-Supervised Pre-training via Masking.}}
\label{sec:pretraining}
To learn generalizable representations, \texttt{TimeRAN} is trained using a masked time-series modeling objective enabled by the \emph{Masking Module}, which is activated only during pre-training. First, a masking ratio $\rho$ is applied to the sequence of patch embeddings produced by the \textit{Projection Module}, where a subset of patches is selected uniformly at random and replaced with a learnable mask token. The partially masked sequence is then processed by the \textit{Time-Series Transformer Encoder} to produce contextualized latent representations. Finally, \texttt{TimeRAN} employs a lightweight reconstruction head that acts as a decoder to reconstruct the input time series from the encoded representations $Z$. It learns accurate representations by minimizing the Mean Squared Error (MSE) between the ground-truth and reconstructed time-series over the masked positions, encouraging robust contextual learning. 

\noindent{\textbf{5.3.6\quad Fine-tuning on RAN Downstream Tasks.}}
\texttt{TimeRAN} supports multiple RAN time-series analytics tasks as mentioned earlier. For anomaly detection and imputation, it leverages the reconstruction head to predict missing values and identify anomalies in the time series. For anomaly detection, the entire input time series is reconstructed, and anomalies are identified based on the reconstruction error aggregated over channels\footnote{Estimating optimal thresholds for anomaly detection is beyond the scope of this study and is left for future work.}. In contrast, for forecasting, \texttt{TimeRAN} replaces the reconstruction head with a forecasting head, which first flattens all the patch embeddings into a \(C \times N \times d\) dimensional vector, and then projects it into a \(C \times H\) dimensional output (values to be forecasted), where \(H\) denotes the forecasting horizon. For classification, \texttt{TimeRAN} uses a classification head that projects the averaged global representation \(z \in \mathbb{R}^{1 \times d}\) into a vector of class logits. Finally, \texttt{TimeRAN} can be fine-tuned under different regimes: (i) end-to-end fine-tuning (\texttt{TimeRAN}$_{\mathrm{FF}}$), (ii) linear probing (\texttt{TimeRAN}$_{\mathrm{LP}}$), where only the task-specific head is trained while the encoder is frozen, and (iii) zero-shot inference (\texttt{TimeRAN}$_{0}$), which is only supported for anomaly detection and imputation, where the pre-trained reconstruction head is used  without additional training. We also explored parameter-efficient fine-tuning using LoRA ~\cite{lora}; however, its performance was not competitive with other regimes and is therefore omitted.

\section{SYSTEM IMPLEMENTATION}
\label{sec:system_implementation}

\noindent\textbf{Prototype Implementation.} We implemented \texttt{TimeRAN} in approximately $\sim$10K lines of Python and Shell code. The system is implemented in Python using the PyTorch and HuggingFace Transformers libraries, while Shell scripting is employed to collect fine-grained runtime resource metrics.

\begin{figure}[t]
\centering
\begin{minipage}{0.42\columnwidth}
    \centering
    \includegraphics[width=\linewidth]{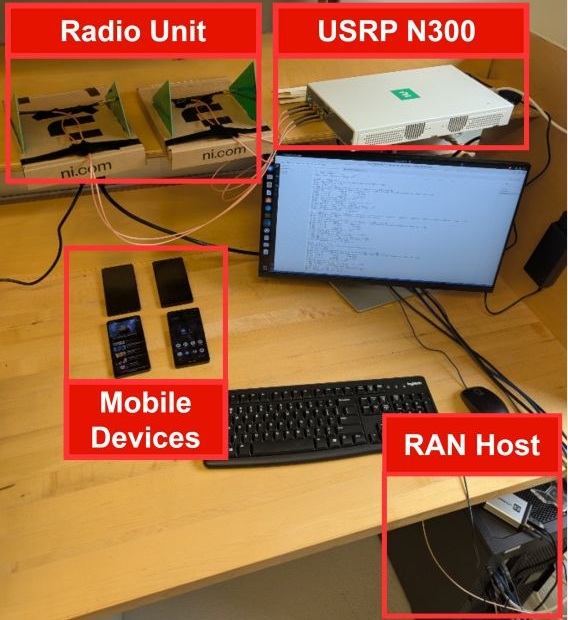}\\
    \vspace{-0.2em}
    \textbf{(a)} 5G RAN.
\end{minipage}\hspace{5mm}%
\begin{minipage}{0.42\columnwidth}
    \centering
    \includegraphics[width=\linewidth]{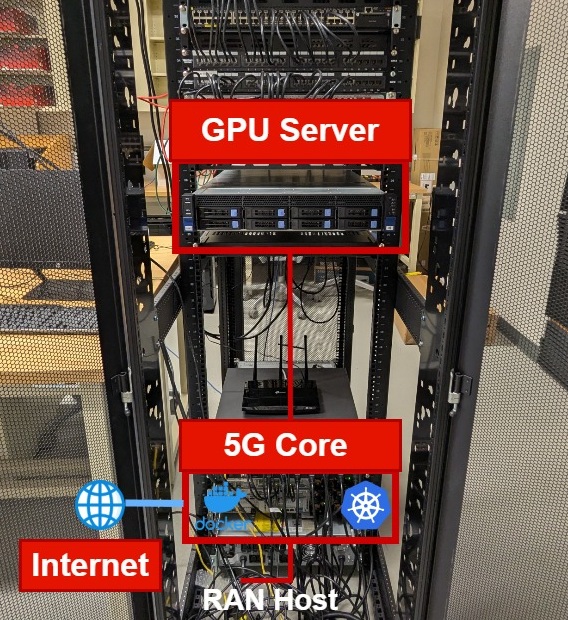}\\
    \vspace{-0.2em}
    \textbf{(b)} 5G Core Network.
\end{minipage}
\vspace{-0.7em}
\caption{Proof-of-concept over-the-air 5G testbed.}
\label{fig:testbed-core}
\vspace{-1.6em}
\end{figure}

\noindent\textbf{System Integration.}
We integrated \texttt{TimeRAN} in a proof-of-concept 5G testbed, as illustrated in Fig.~\ref{fig:testbed-core}. The setup consists of a disaggregated 5G base station (Fig.~\ref{fig:testbed-core}(a)) coupled with a Software-Defined Radio (SDR) frontend (USRP N300), a cloud-native 5G core (Fig.~\ref{fig:testbed-core}(b)), and four Google Pixel 7 mobile devices enabling fully over-the-air experimentation. The RAN baseband functions run on a high-performance computing platform equipped with an AMD Ryzen Threadripper PRO 5975WX processor (32 cores) and 504~GB of DDR4 memory. GPU acceleration is provided by separate GPU servers equipped with NVIDIA RTX 3090, RTX 4090, and RTX A6000 GPUs. The core network runs on a separate server with an AMD EPYC 7352 24-core processor and 128~GB of DDR4 memory. The base station connects to the SDR via a dedicated 10~Gbps fronthaul link, with the SDR interfaced to 6~dBi directional antennas, and communicates with the core through a 10~Gbps backhaul link. Both the RAN and core network are implemented using the OAI software stack~\cite{oai}; the core network follows a cloud-native microservice-based architecture where all the virtual network functions, including the AMF, SMF, NRF, UPF, UDM, UDR, and AUSF, are deployed as containerized pods in an on-premises Kubernetes cluster. The system operates in the n78 TDD band with 20~MHz bandwidth centered at 3.319~GHz and follows a 5~ms radio frame structure configured with seven downlink slots and two uplink slots with 30~kHz subcarrier spacing.

\section{EVALUATION SETUP}

\subsection{Corpus Partitioning and Benchmarking}
\label{sec:benchmark}
\noindent We construct disjoint training and test splits, for each dataset in the \textit{TimeRAN DataPile}. When official splits are provided by the dataset creators, we adopt them directly; otherwise, we apply a 70\%/30\% train--test split strategy. For long continuous time-series datasets, we perform a temporal split, assigning the first 70\% of the sequence to training and the remaining 30\% to testing. For datasets consisting of multiple independent short time-series, we assign each complete time series exclusively to either the training or the test set.

During pre-training, we restrict \texttt{TimeRAN} to the training splits of each dataset, excluding both test splits and anomaly detection datasets to prevent information leakage and ensure that only normal temporal patterns are learned. For evaluation, we also construct a comprehensive and diverse benchmark using a representative subset of the \textit{TimeRAN DataPile} datasets. As illustrated in Table~\ref{tab:benchmark}, five datasets are selected for each downstream task to capture diverse telemetry characteristics, temporal granularities, and operational conditions across the RAN protocol stack.

% To ensure reproducibility, all experiments are conducted using a fixed random seed of 77.

\begin{table}[t]
\vspace{-1em}
\centering
\scriptsize
\renewcommand{\arraystretch}{1.1}
\setlength{\tabcolsep}{3pt}

\begin{tabularx}{\columnwidth}{
>{\centering\arraybackslash}p{0.23\columnwidth}
>{\centering\arraybackslash}p{0.27\columnwidth}
>{\centering\arraybackslash}p{0.45\columnwidth}
}
\toprule
\textbf{Tasks} & \textbf{Supervision} & \textbf{Datasets} \\
\midrule

Anomaly Detection
& \makecell{Zero-shot \\ Linear probing}
& \makecell{AERPAW 18/23 \cite{aerpaw18_dataset,aerpaw23_dataset}, BLT \cite{blt_dataset}, \\ Hetnets \cite{hetnets_dataset}, Jamshield \cite{jamshield_dataset}} \\[6pt]

Classification
& \makecell{Linear probing \\ Full-finetuning}
& \vspace{-10pt} AERPAW 20 \cite{aerpaw20_dataset}, Irish \cite{irish_dataset}, Spotlight \cite{spotlight_dataset}, TelecomTS \cite{telecomts_dataset}, Tractor \cite{tractor_dataset} \\[6pt]

Forecasting
& \makecell{Linear probing \\ Full-finetuning}
& \vspace{-10pt} AERPAW 18/24/25 \cite{aerpaw18_dataset, aerpaw24_dataset, aerpaw25_dataset}, QoE Aware \cite{qoe_aware_dataset}, Queens \cite{queens_dataset} \\[6pt]

Imputation
& \makecell{Zero-shot \\ Linear probing}
& \makecell{Irish \cite{irish_dataset}, Open Ireland \cite{open_ireland_dataset}, \\ Hetnets \cite{hetnets_dataset}, WINS \cite{wins_dataset}, NOK \cite{nok_dataset}} \\

\bottomrule
\end{tabularx}
\caption{TimeRAN evaluation benchmark.}
\label{tab:benchmark}
\vspace{-2em}
\end{table}

\subsection{Configuration and Training Settings}

\noindent\textbf{TimeRAN Configuration.}
We leverage the pretrained encoder of MOMENT \cite{moment} as the backbone of our \textit{Time-Series Transformer Encoder}, due to its state-of-the-art performance achieved through training on billions of time-series samples across diverse domains, as well as its lightweight architecture, which enables efficient learning. We then perform continued pre-training on the \textit{TimeRAN DataPile} to adapt the model to RAN telemetry data. 
We train three \texttt{TimeRAN} variants with capacities aligned to the  sizes of the T5 encoder \cite{t5_transformer}. Specifically, the $\texttt{Base}$ ($\underline{\texttt{Small}}$, $\overline{\texttt{Large}}$) configuration employs a transformer encoder with $L = 12$ ($\underline{6}$, $\overline{24}$) layers, embedding dimension $d = 768$ ($\underline{512}$, $\overline{1024}$), $B = 12$ ($\underline{8}$, $\overline{16}$) attention heads, and feed-forward dimensions of $3072$ ($\underline{2048}$, $\overline{4096}$), resulting in approximately $125$ ($\underline{40}$, $\overline{385}$) million parameters. Each variant processes multivariate time-series inputs with observation window length of $T = 512$, which are then partitioned into $N = 64$ non-overlapping patches of length $P = 8$. Prior to projection, patches are normalized using reversible instance normalization. Sinusoidal absolute positional encodings are added to the patch embeddings to preserve temporal ordering across the sequence. During continued pre-training, $\rho = 30\%$ of the patch embeddings are randomly masked following a uniform sampling strategy, enabling self-supervised learning over the input sequence. Finally, all task-specific heads employ $M = 2$ linear layers.

\begin{figure}[t]
\centering
\begin{subfigure}[t]{0.48\columnwidth}
    \centering
    \includegraphics[width=\linewidth]{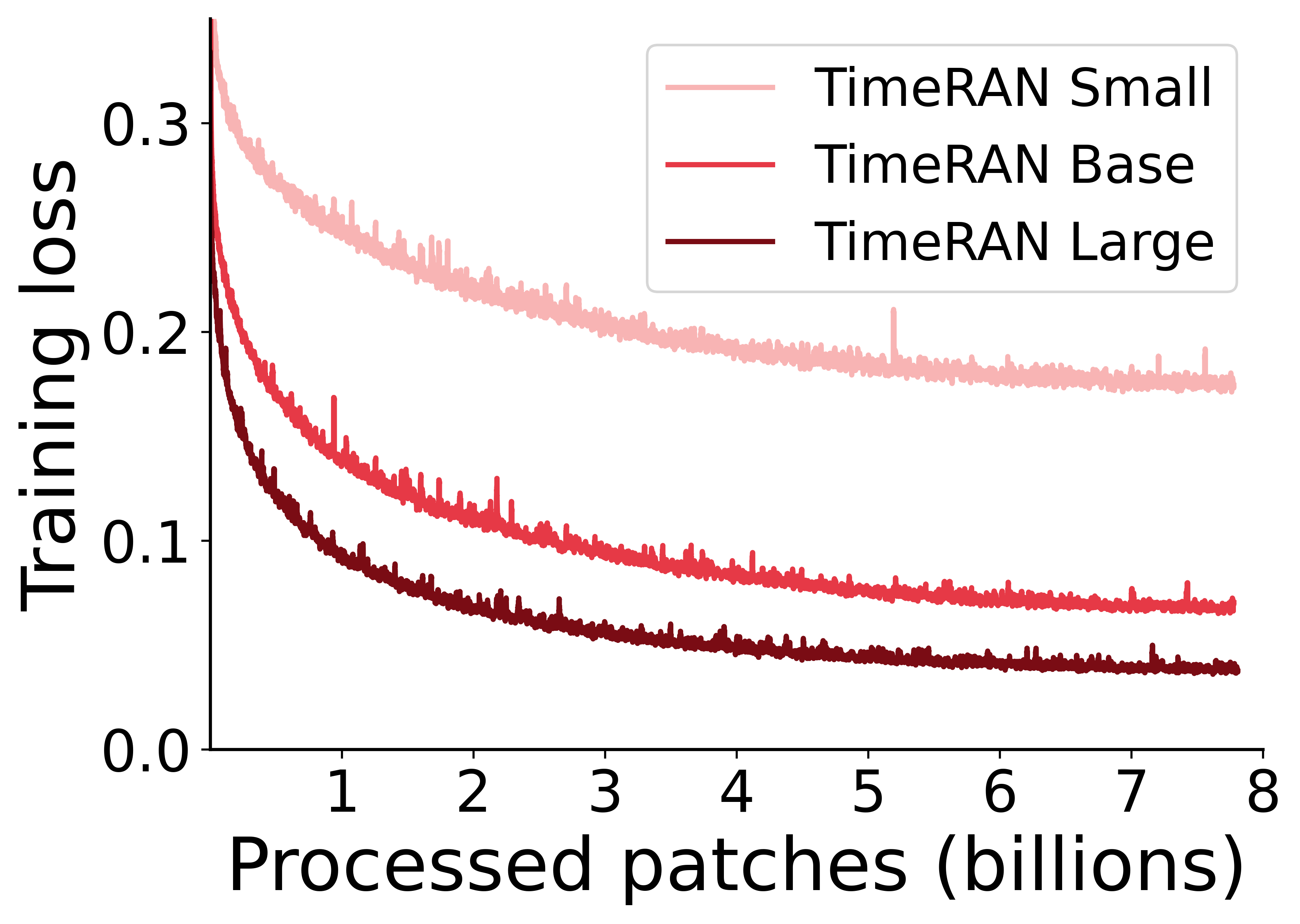}
    \vspace{-1.5em}
    \caption{Training loss.}
    \label{fig:loss}
\end{subfigure}
\hfill
\begin{subfigure}[t]{0.48\columnwidth}
    \centering
    \includegraphics[width=\linewidth]{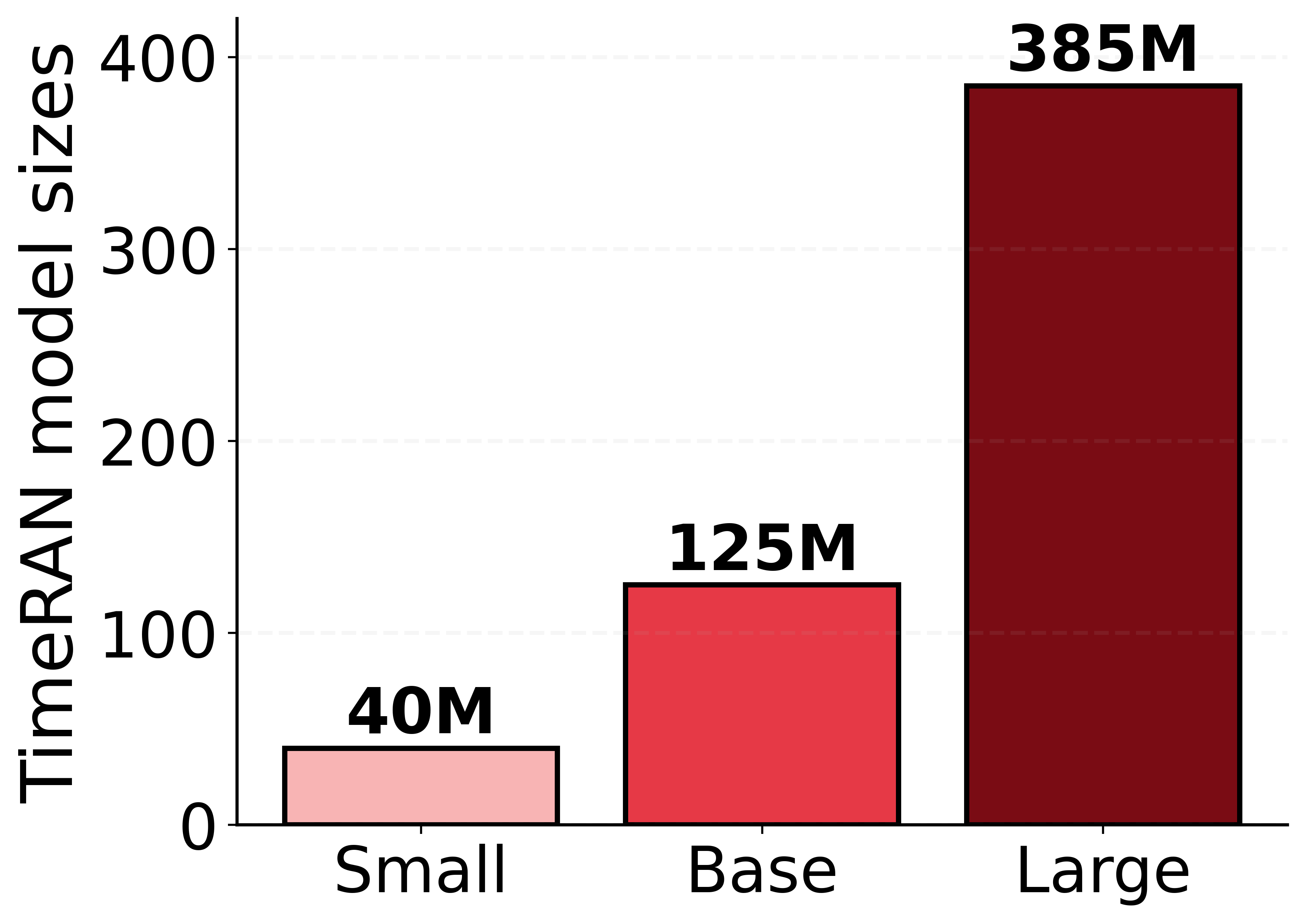}
    \vspace{-1.5em}
    \caption{Number of parameters.}
    \label{fig:sizes}
\end{subfigure}
\vspace{-1em}
\caption{Training loss and size of TimeRAN variants.}
\label{fig:training_summary}
\vspace{-1em}
\end{figure}

\noindent\textbf{Training Setup.}
\texttt{TimeRAN} variants are trained using the Adam optimizer with decoupled weight decay, with $\lambda = 0.05$, $\beta_1 = 0.9$, and $\beta_2 = 0.999$. Gradients are clipped at a maximum norm of $5.0$, and training is performed with a batch size of $2048$. We employ a cosine learning rate schedule with initial and final learning rates of $10^{-4}$ and $10^{-5}$, respectively. 
During pre-training, we improve sample efficiency via overlapping patching with stride values $\{1,2,4\}$, selected based on dataset size, with smaller datasets using finer strides and larger datasets using coarser strides. Dataset heterogeneity is addressed through a size-aware sampling strategy that inversely reweights datasets, preventing dominance of large datasets during continued pre-training.  Training efficiency and memory usage are further optimized through gradient checkpointing and mixed-precision training. All model variants are trained for one full epoch in a mixed-precision setting, using \texttt{float32} for numerically unstable operations and \texttt{bfloat16} for all other computations. Fig.~\ref{fig:training_summary} illustrates the convergence of the training loss of the \texttt{TimeRAN} variants.

\noindent\textbf{Fine-tuning Setup.} Unless stated otherwise, we employ \texttt{TimeRAN-Base} (approximately 125M parameters, requiring less than \texttt{500 MB} in \texttt{float32}), which provides a favorable balance between performance, training efficiency, and resource footprint, making it well-suited for deployments with limited compute and memory resources. \texttt{TimeRAN-Large} achieves higher performance, as illustrated in Fig.~\ref{fig:training_summary}, though full results are omitted due to space limitations. Finally, detailed configurations of the fine-tuning strategies and task-specific hyperparameters are provided in~\cite{timeran_datapile}.

\subsection{Baselines}
\noindent We compare \texttt{TimeRAN} with state-of-the-art deep learning and statistical models across tasks. These include pretrained general-purpose time-series foundation models (e.g., MOMENT~\cite{moment}), transformer-based time-series architectures (e.g., Autoformer~\cite{autoformer}, Informer~\cite{informer}, TimesNet~\cite{timesnet}), deep learning sequential models~\cite{deep_learning_baselines} (e.g., 1D-CNN, LSTM), and classical statistical methods~\cite{arima} (e.g., ARIMA, ETS). Notably, LSTM- and CNN-based approaches are widely employed for time-series tasks in the telecommunications domain, and we adopt architectures consistent with those reported in prior work~\cite{jamshield_dataset, tractor_dataset, hetnets_dataset}. For a fair comparison, we use model configurations with parameter sizes comparable to \texttt{TimeRAN-Base} for the transformer architectures. Detailed architectural and hyperparameter settings for all methods are provided in~\cite{timeran_datapile}.

\subsection{Evaluation Metrics}
\noindent We assess \texttt{TimeRAN} using standard task-specific metrics commonly adopted in the time-series literature. For anomaly detection, we use the Adjusted Best F1 score~\cite{adjusted_f1}, which selects the optimal threshold over reconstruction-based anomaly scores (e.g., MSE) to maximize the F1 score, where an anomaly segment is considered correctly detected if any timestep within the segment is identified. For classification, we report Precision (P), Recall (R), and F1-score. For imputation and forecasting, we use MSE and Mean Absolute Error (MAE) to evaluate prediction accuracy.

\section{EVALUATION RESULTS}
\label{sec:evaluation_results}

\noindent Our evaluation is guided by the following research questions:

\noindent\textbf{RQ1:}
How well does \texttt{TimeRAN} perform across diverse RAN downstream tasks (\S\ref{sec:task_performance})?

\noindent\textbf{RQ2:}
To what extent does \texttt{TimeRAN} learn generalizable and transferable representations across heterogeneous datasets and previously unseen environments (\S\ref{sec:generalization_analysis})?

\noindent\textbf{RQ3:}
What resource overheads does \texttt{TimeRAN} incur  (\S\ref{sec:system_overhead})?

\noindent\textbf{RQ4:}
How does \texttt{TimeRAN} perform in real-world scenarios across multiple downstream tasks (\S\ref{sec:integration})?

\subsection{Downstream Task Performance}
\label{sec:task_performance}

\noindent To answer RQ1, we evaluate \texttt{TimeRAN} across multiple downstream tasks using the benchmark described in Section~\S\ref{sec:benchmark}. As the selected datasets in the curated benchmark capture different target variables and scales, we refer the reader to the corresponding dataset references for further details.

\noindent{\textbf{8.1.1\quad Anomaly Detection.}} \texttt{TimeRAN} consistently outperforms all baselines in anomaly detection (Table~\ref{tab:anomaly_detection_results}). In particular, \texttt{TimeRAN}$_{\mathrm{LP}}$ achieves the strongest performance, indicating that representations learned from normal RAN time series enable accurate anomaly detection without full fine-tuning. Furthermore, \texttt{TimeRAN} remains effective in the zero-shot setting, surpassing most methods, with consistent gains across both univariate and multivariate time series datasets.

\noindent{\textbf{8.1.2\quad Classification.}} For classification, \texttt{TimeRAN} with full fine-tuning achieves state-of-the-art performance across representative tasks (Table~\ref{tab:classification_results}). Moreover, deep learning sequential models often outperform transformer-based time-series baselines under limited labeled data due to stronger inductive biases and more data-efficient training. Nevertheless, \texttt{TimeRAN} remains highly adaptable and consistently matches or exceeds their performance when fully fine-tuned.

\noindent{\textbf{8.1.3\quad Forecasting.}} In forecasting, \texttt{TimeRAN} achieves satisfactory performance across datasets and horizons (Table~\ref{tab:forecasting_results}). While no single method dominates, \texttt{TimeRAN} remains competitive without extensive fine-tuning. Performance degrades with increasing forecasting horizon for all methods, reflecting the inherent uncertainty of long-range prediction. In this regime, \texttt{TimeRAN}$_{\mathrm{LP}}$ outperforms both fully fine-tuned \texttt{TimeRAN} and MOMENT variants, while deep learning approaches consistently outperform statistical methods due to the high dimensionality and stochasticity of RAN telemetry.

\noindent{\textbf{8.1.4\quad Imputation.}} Finally, \texttt{TimeRAN} achieves notable imputation performance across masking ratios (Table~\ref{tab:imputation_results}). As the masking ratio increases, reconstruction becomes more challenging, leading to degradation across all methods. Nevertheless, \texttt{TimeRAN} remains robust under both training settings, consistently outperforming most statistical methods.

\noindent Overall, the results indicate that \texttt{TimeRAN} provides a unified and effective solution across diverse RAN downstream analytics tasks, achieving strong performance under both minimal adaptation and full fine-tuning strategies. 

\begin{table*}[t]
\centering
\scriptsize
\setlength{\tabcolsep}{3.0pt}
\renewcommand{\arraystretch}{1.12}

\begin{tabular}{c| c| c| c| c| c| c| c}
\hline
\multicolumn{1}{c|}{\textbf{Methods}} &
\multicolumn{1}{c|}{\textbf{TimeRAN$_\mathrm{LP}$}} &
\multicolumn{1}{c|}{\textbf{TimeRAN$_\mathrm{0}$}} &
\multicolumn{1}{c|}{\textbf{MOMENT$_\mathrm{LP}$}} &
\multicolumn{1}{c|}{\textbf{MOMENT$_\mathrm{0}$}} &
\multicolumn{1}{c|}{\textbf{Autoformer}} &
\multicolumn{1}{c|}{\textbf{Informer}} &
\multicolumn{1}{c}{\textbf{TimesNet}} \\

\textbf{Dataset} &
\textbf{Adj.\ F1} &
\textbf{Adj.\ F1} &
\textbf{Adj.\ F1} &
\textbf{Adj.\ F1} &
\textbf{Adj.\ F1} &
\textbf{Adj.\ F1} &
\textbf{Adj.\ F1} \\
\hline

\textbf{AERPAW 18} \cite{aerpaw18_dataset}
& \textbf{0.93} & 0.86 & 0.83 & 0.90 & 0.91 & 0.90 & 0.89 \\
\hline

\textbf{AERPAW 23} \cite{aerpaw23_dataset}
& \textbf{0.99} & 0.98 & 0.98 & 0.98 & 0.93 & 0.94 & 0.94 \\
\hline

\textbf{BLT} \cite{blt_dataset}
& \textbf{0.98} & 0.95 & 0.93 & 0.91 & 0.86 & 0.89 & 0.83 \\
\hline

\textbf{Hetnets} \cite{hetnets_dataset}
& \textbf{0.99} & 0.99 & 0.98 & 0.98 & 0.96 & 0.95 & 0.96 \\
\hline

\textbf{Jamshield} \cite{jamshield_dataset}
& \textbf{0.97} & 0.93 & 0.91 & 0.90 & 0.92 & 0.92 & 0.93 \\
\hline

\end{tabular}
\caption{Anomaly detection performance on the curated evaluation benchmark.}
\vspace{-1em}
\label{tab:anomaly_detection_results}
\end{table*}

\begin{table*}[t]
\centering
\scriptsize
\setlength{\tabcolsep}{2.8pt}
\renewcommand{\arraystretch}{1.1}

\begin{tabular}{l c| ccc| ccc| ccc| ccc| ccc| ccc| ccc}
\hline
\multicolumn{2}{c|}{\textbf{Methods}} &
\multicolumn{3}{c|}{\textbf{TimeRAN}$_{\textbf{FF}}$} &
\multicolumn{3}{c|}{\textbf{TimeRAN}$_{\textbf{LP}}$} &
\multicolumn{3}{c|}{\textbf{MOMENT}$_{\textbf{FF}}$} &
\multicolumn{3}{c|}{\textbf{Informer}} &
\multicolumn{3}{c|}{\textbf{TimesNet}} &
\multicolumn{3}{c|}{\textbf{1D-CNN}} &
\multicolumn{3}{c}{\textbf{LSTM}} \\

\textbf{Dataset} & \textbf{Task} &
\textbf{P.} & \textbf{R.} & \textbf{F1} &
\textbf{P.} & \textbf{R.} & \textbf{F1} &
\textbf{P.} & \textbf{R.} & \textbf{F1} &
\textbf{P.} & \textbf{R.} & \textbf{F1} &
\textbf{P.} & \textbf{R.} & \textbf{F1} &
\textbf{P.} & \textbf{R.} & \textbf{F1} &
\textbf{P.} & \textbf{R.} & \textbf{F1}\\
\hline

\textbf{AERPAW 20} \cite{aerpaw20_dataset} & Location
& \textbf{1.00} & \textbf{0.99} & \textbf{0.99}
& 0.62 & 0.80 & 0.70
& 1.00 & 0.98 & 0.99
& 0.65 & 0.76 & 0.71
& 0.61 & 0.69 & 0.68
& 0.95 & 0.90 & 0.93
& 0.68 & 0.72 & 0.71 \\
\hline

\textbf{Irish}\cite{irish_dataset} & Mobility
& \textbf{1.00} & \textbf{0.99} & \textbf{0.99}
& 0.96 & 0.89 & 0.92
& 0.99 & 0.99 & 0.99
& 0.95 & 0.91 & 0.92
& 0.96 & 0.92 & 0.94
& 0.98 & 0.99 & 0.98
& 0.96 & 0.93 & 0.95 \\
\hline

\multirow{2}{*}{\textbf{Spotlight} \cite{spotlight_dataset}} 
& Anomaly
& 0.93 & 0.94 & 0.93
& 0.84 & 0.83 & 0.84
& \textbf{0.93} & \textbf{0.95} & \textbf{0.94}
& 0.83 & 0.89 & 0.86
& 0.81 & 0.89 & 0.84
& 0.78 & 0.92 & 0.81
& 0.90 & 0.83 & 0.87 \\
& Root-Cause
& \textbf{0.99} & \textbf{0.99} & \textbf{0.99}
& 0.55 & 0.45 & 0.40
& 0.96 & 0.95 & 0.95
& 0.57 & 0.58 & 0.58
& 0.61 & 0.59 & 0.57
& 0.63 & 0.66 & 0.61
& 0.64 & 0.65 & 0.64 \\
\hline

\multirow{2}{*}{\textbf{TelecomTS} \cite{telecomts_dataset}} 
& Congestion
& \textbf{0.98} & \textbf{1.00} & \textbf{0.99}
& 0.58 & 0.95 & 0.72
& 0.97 & 0.98 & 0.98
& 0.74 & 0.99 & 0.85
& 0.72 & 0.99 & 0.84
& 0.87 & 0.93 & 0.90
& 0.75 & 0.99 & 0.86 \\
& Coverage
& 0.88 & 0.89 & 0.88
& 0.54 & 0.55 & 0.55
& 0.81 & 0.81 & 0.80
& 0.80 & 0.81 & 0.79
& 0.82 & 0.81 & 0.82
& 0.81 & 0.80 & 0.81
& \textbf{0.90} & \textbf{0.89} & \textbf{0.89} \\
\hline

\multirow{2}{*}{\textbf{Tractor} \cite{tractor_dataset}} 
& Services
& 0.97 & 0.97 & 0.97
& 0.75 & 0.72 & 0.71
& \textbf{0.98} & \textbf{0.98} & \textbf{0.98}
& 0.70 & 0.68 & 0.68
& 0.70 & 0.62 & 0.61
& 0.90 & 0.88 & 0.87
& 0.81 & 0.79 & 0.79 \\
& Slice Type
& \textbf{0.99} & \textbf{0.99} & \textbf{0.99}
& 0.83 & 0.81 & 0.80
& 0.98 & 0.98 & 0.98
& 0.92 & 0.91 & 0.91
& 0.89 & 0.90 & 0.90
& 0.96 & 0.97 & 0.96
& 0.94 & 0.94 & 0.94 \\
\hline

\end{tabular}
\caption{Classification performance on the curated evaluation benchmark.}
\vspace{-1em}
\label{tab:classification_results}
\end{table*}

\begin{table*}[t]
\centering
\scriptsize
\setlength{\tabcolsep}{3.2pt}
\renewcommand{\arraystretch}{1.05}

\resizebox{\textwidth}{!}{%
\begin{tabular}{l c| cc| cc| cc| cc| cc| cc| cc| cc| cc| cc}
\hline
\multicolumn{2}{c|}{\textbf{Methods}} &
\multicolumn{2}{c|}{\textbf{TimeRAN}$_{\textbf{FF}}$} &
\multicolumn{2}{c|}{\textbf{TimeRAN}$_{\textbf{LP}}$} &
\multicolumn{2}{c|}{\textbf{MOMENT}$_{\textbf{FF}}$} &
\multicolumn{2}{c|}{\textbf{Autoformer}} &
\multicolumn{2}{c|}{\textbf{Informer}} &
\multicolumn{2}{c|}{\textbf{TimesNet}} &
\multicolumn{2}{c|}{\textbf{1D-CNN}} &
\multicolumn{2}{c|}{\textbf{LSTM}} &
\multicolumn{2}{c|}{\textbf{ETS}} &
\multicolumn{2}{c}{\textbf{ARIMA}} \\

\textbf{Dataset} & \textbf{Horizon (H)} &
\textbf{MSE} & \textbf{MAE}
& \textbf{MSE} & \textbf{MAE}
& \textbf{MSE} & \textbf{MAE}
& \textbf{MSE} & \textbf{MAE}
& \textbf{MSE} & \textbf{MAE}
& \textbf{MSE} & \textbf{MAE}
& \textbf{MSE} & \textbf{MAE}
& \textbf{MSE} & \textbf{MAE}
& \textbf{MSE} & \textbf{MAE}
& \textbf{MSE} & \textbf{MAE} \\
\hline

\multirow{4}{*}{\textbf{AERPAW 18} \cite{aerpaw18_dataset}} & 32
& 25.55 & 3.86 & 23.55 & 3.78 & 31.92 & 4.33 & 21.70 & 3.55 & 31.30 & 4.16 & 22.78 & 3.67 & 28.63 & 4.08 & \textbf{20.43} & \textbf{3.40} & 34.32 & 4.61 & 31.78 & 4.33 \\
& 64
& 26.46 & 3.99 & 26.29 & 3.97 & 31.61 & 4.32 & \textbf{22.79} & \textbf{3.66} & 23.47 & 3.77 & 24.46 & 3.82 & 27.41 & 4.05 & 22.92 & 3.70 & 38.81 & 4.78 & 34.12 & 4.60 \\
& 128
& 28.17 & 4.07 & 25.15 & 3.86 & 27.45 & 4.05 & 23.36 & 3.76 & 24.63 & 3.84 & 24.11 & 3.81 & 25.66 & 3.93 & \textbf{23.26} & \textbf{3.73} & 39.21 & 4.81 & 37.21 & 4.71 \\
& 208
& 28.69 & 4.10 & 25.71 & 3.93 & 27.32 & 4.04 & 25.25 & 3.87 & 25.81 & 3.96 & \textbf{24.40} & \textbf{3.81} & 24.69 & 3.85 & 24.84 & 3.87 & 41.44 & 5.11 & 40.11 & 5.01 \\
\hline

\multirow{4}{*}{\textbf{AERPAW 24} \cite{aerpaw24_dataset}} & 32
& 17.70 & 3.53 & 14.75 & 3.14 & 19.00 & 3.56 & 16.86 & 3.50 & 20.26 & 3.79 & \textbf{14.27} & \textbf{3.10} & 27.33 & 4.39 & 15.45 & 3.18 & 29.80 & 4.65 & 26.48 & 4.32 \\
& 64
& 17.72 & 3.53 & 14.67 & 3.13 & 20.17 & 3.77 & 23.77 & 3.97 & 16.48 & 3.23 & \textbf{14.19} & \textbf{3.09} & 27.36 & 4.40 & 16.65 & 3.43 & 44.94 & 5.13 & 44.11 & 5.10 \\
& 128
& 19.67 & 3.74 & 17.02 & 3.51 & \textbf{14.11} & \textbf{3.07} & 23.37 & 3.95 & 16.87 & 3.50 & 19.13 & 3.71 & 25.58 & 4.11 & 19.01 & 3.69 & 68.26 & 6.35 & 69.90 & 6.41 \\
& 208
& 36.79 & 5.07 & 27.08 & 4.37 & \textbf{13.99} & \textbf{2.96} & 25.74 & 4.27 & 19.76 & 3.74 & 22.72 & 3.85 & 29.60 & 4.61 & 30.90 & 4.84 & 110.22 & 8.21 & 109.92 & 8.13 \\
\hline

\multirow{4}{*}{\textbf{AERPAW 25} \cite{aerpaw25_dataset}} & 32
& 6.35 & 2.06 & 5.79 & 1.93 & 6.25 & 2.05 & 5.27 & 1.86 & 5.25 & 1.85 & 5.32 & 1.86 & 5.92 & 1.95 & \textbf{5.18} & \textbf{1.83} & 5.96 & 2.01 & 6.00 & 2.01 \\
& 64
& 6.56 & 2.07 & 5.85 & 1.94 & 6.43 & 2.06 & 5.57 & 1.90 & 5.27 & 1.86 & 5.31 & 1.86 & 5.81 & 1.93 & \textbf{5.23} & \textbf{1.85} & 6.01 & 2.02 & 6.08 & 2.02 \\
& 128
& 6.83 & 2.08 & 5.89 & 1.95 & 6.56 & 2.07 & 5.70 & 1.92 & 5.29 & 1.86 & 5.34 & 1.87 & 5.83 & 1.94 & \textbf{5.28} & \textbf{1.86} & 6.08 & 2.03 & 6.17 & 2.04 \\
& 208
& 6.85 & 2.08 & 5.90 & 1.95 & 6.53 & 2.07 & 5.52 & 1.90 & 5.29 & 1.86 & 5.31 & 1.86 & 5.90 & 1.95 & \textbf{5.29} & \textbf{1.86} & 6.12 & 2.03 & 6.23 & 2.05 \\
\hline

\multirow{4}{*}{\textbf{QoE Aware} \cite{qoe_aware_dataset}} & 32
& 48.81 & 4.98 & \textbf{46.93} & \textbf{4.91} & 51.03 & 5.15 & 52.15 & 5.21 & 51.05 & 5.15 & 50.88 & 5.14 & 61.06 & 5.64 & 47.13 & 4.93 & 71.63 & 6.01 & 54.68 & 5.31 \\
& 64
& 56.23 & 5.42 & \textbf{50.05} & \textbf{5.10} & 56.98 & 5.41 & 53.21 & 5.26 & 52.48 & 5.22 & 52.42 & 5.22 & 62.36 & 5.70 & 50.68 & 5.11 & 84.51 & 6.41 & 62.96 & 5.72 \\
& 128
& 60.17 & 5.60 & \textbf{52.63} & \textbf{5.24} & 60.50 & 5.59 & 54.46 & 5.31 & 54.18 & 5.30 & 54.22 & 5.31 & 63.70 & 5.76 & 53.32 & 5.26 & 95.13 & 6.90 & 68.68 & 5.88 \\
& 208
& 62.27 & 5.68 & \textbf{54.18} & \textbf{5.31} & 62.85 & 5.71 & 55.16 & 5.35 & 55.12 & 5.34 & 55.24 & 5.35 & 64.60 & 5.80 & 54.82 & 5.33 & 101.24 & 7.16 & 73.02 & 6.07 \\
\hline

\multirow{4}{*}{\textbf{Queens} \cite{queens_dataset}} & 32
& 17.29 & 3.03 & 17.88 & 3.14 & \textbf{16.62} & \textbf{2.88} & 46.75 & 5.05 & 27.64 & 3.91 & 27.04 & 3.86 & 46.77 & 5.09 & 22.20 & 3.40 & 17.94 & 3.18 & 16.98 & 2.93 \\
& 64
& 24.76 & 3.64 & 23.89 & 3.63 & \textbf{23.63} & \textbf{3.42} & 50.32 & 5.21 & 35.39 & 4.38 & 36.18 & 4.43 & 51.77 & 5.36 & 29.34 & 3.94 & 26.50 & 3.76 & 26.03 & 3.66 \\
& 128
& 35.91 & 4.42 & \textbf{32.29} & \textbf{4.24} & 33.45 & 4.30 & 55.46 & 5.55 & 44.97 & 4.99 & 45.83 & 5.03 & 59.50 & 5.75 & 40.39 & 4.47 & 42.53 & 4.90 & 42.32 & 4.49 \\
& 208
& 48.32 & 5.17 & \textbf{42.47} & \textbf{4.86} & 44.39 & 4.92 & 60.57 & 5.79 & 53.65 & 5.49 & 53.46 & 5.46 & 63.78 & 5.95 & 52.10 & 5.40 & 64.48 & 5.99 & 64.20 & 5.98 \\
\hline

\end{tabular}
}
\caption{Forecasting performance on the curated evaluation benchmark.}
\vspace{-1em}
\label{tab:forecasting_results}
\end{table*}

\begin{table*}[t]
\centering
\scriptsize
\setlength{\tabcolsep}{3.0pt}
\renewcommand{\arraystretch}{1.12}

\begin{tabular}{c c| *{7}{c c|} c c}
\hline
\multicolumn{2}{c|}{\textbf{Methods}} &
\multicolumn{2}{c|}{\textbf{TimeRAN$_\mathrm{LP}$}} &
\multicolumn{2}{c|}{\textbf{TimeRAN$_\mathrm{0}$}} &
\multicolumn{2}{c|}{\textbf{MOMENT$_\mathrm{FF}$}} &
\multicolumn{2}{c|}{\textbf{Forward Fill}} &
\multicolumn{2}{c|}{\textbf{Mean}} &
\multicolumn{2}{c|}{\textbf{N. Neighbors}} &
\multicolumn{2}{c|}{\textbf{Linear}} &
\multicolumn{2}{c}{\textbf{Rolling Mean}} \\

\textbf{Dataset} & \textbf{Mask Ratio} &
\textbf{MSE} & \textbf{MAE} &
\textbf{MSE} & \textbf{MAE} &
\textbf{MSE} & \textbf{MAE} &
\textbf{MSE} & \textbf{MAE} &
\textbf{MSE} & \textbf{MAE} &
\textbf{MSE} & \textbf{MAE} &
\textbf{MSE} & \textbf{MAE} &
\textbf{MSE} & \textbf{MAE} \\
\hline

% ---------------- Irish ----------------
\multirow[c]{4}{*}{\centering\textbf{Irish} \cite{irish_dataset}} & 10\%
& \textbf{4.95} & \textbf{1.44}
& 5.16 & 1.48
& 23.04 & 3.02
& 13.02 & 2.34
& 20.82 & 2.92
& 10.80 & 1.94
& 8.45 & 1.86
& 12.18 & 2.26 \\
& 30\%
& \textbf{5.96} & \textbf{1.56}
& 6.21 & 1.60
& 22.55 & 2.98
& 16.72 & 2.53
& 24.59 & 3.07
& 14.50 & 2.40
& 11.86 & 2.00
& 14.68 & 2.42 \\
& 50\%
& \textbf{7.38} & \textbf{1.75}
& 8.47 & 1.87
& 24.40 & 3.08
& 16.77 & 2.57
& 23.67 & 3.05
& 15.56 & 2.49
& 12.40 & 2.28
& 14.90 & 2.45 \\
\hline

% ---------------- Open Ireland ----------------
\multirow[c]{4}{*}{\centering\textbf{Open Ireland} \cite{open_ireland_dataset}} & 10\%
& 21.02 & 2.54
& \textbf{20.80} & \textbf{2.53}
& 23.64 & 2.78
& 62.46 & 4.38
& 34.52 & 3.78
& 62.33 & 4.37
& 52.35 & 4.17
& 34.46 & 3.61 \\
& 30\%
& 22.75 & 2.69
& \textbf{22.67} & \textbf{2.68}
& 24.60 & 2.87
& 62.80 & 4.40
& 34.90 & 3.80
& 62.55 & 4.38
& 52.16 & 4.16
& 35.03 & 3.80 \\
& 50\%
& \textbf{24.99} & \textbf{2.88}
& 25.04 & 2.90
& 26.53 & 3.03
& 63.79 & 4.45
& 35.04 & 3.81
& 63.21 & 4.41
& 52.90 & 4.18
& 35.82 & 3.84 \\
\hline

% ---------------- Hetnets ----------------
\multirow[c]{4}{*}{\centering\textbf{Hetnets} \cite{hetnets_dataset}} & 10\%
& \textbf{0.58} & \textbf{0.56}
& 0.67 & 0.59
& 3.42 & 1.17
& 7.21 & 1.65
& 8.84 & 2.29
& 5.64 & 1.38
& 4.40 & 1.31
& 6.01 & 1.47 \\
& 30\%
& 0.78 & 0.61
& \textbf{0.76} & \textbf{0.60}
& 4.13 & 1.29
& 7.70 & 1.67
& 9.38 & 2.32
& 6.22 & 1.50
& 5.06 & 1.36
& 6.27 & 1.53 \\
& 50\%
& 1.03 & 0.65
& \textbf{1.01} & \textbf{0.64}
& 4.55 & 1.34
& 8.29 & 1.71
& 9.12 & 2.31
& 7.07 & 1.62
& 5.70 & 1.40
& 6.52 & 1.58 \\
\hline

% ---------------- WINS ----------------
\multirow[c]{4}{*}{\centering\textbf{WINS} \cite{wins_dataset}} & 10\%
& 246.03 & 5.88
& \textbf{144.70} & \textbf{5.22}
& 280.53 & 6.37
& 1669.83 & 16.87
& 8094.32 & 48.35
& 701.80 & 10.68
& 442.69 & 7.62
& 2712.92 & 24.01 \\
& 30\%
& 534.79 & 8.78
& \textbf{523.31} & \textbf{8.75}
& 597.65 & 9.21
& 2268.78 & 20.24
& 7883.61 & 47.87
& 1175.44 & 13.83
& 864.97 & 11.24
& 3172.77 & 26.68 \\
& 50\%
& \textbf{1104.23} & \textbf{12.98}
& 1122.85 & 13.19
& 1182.96 & 13.86
& 3365.68 & 27.11
& 8004.40 & 48.02
& 2072.07 & 18.72
& 1615.33 & 16.27
& 3985.17 & 30.01 \\
\hline

% ---------------- NOK ----------------
\multirow[c]{4}{*}{\centering\textbf{NOK} \cite{nok_dataset}} & 10\%
& 6.10 & 0.69
& 5.78 & 0.66
& 8.72 & 0.86
& 9.15 & 0.93
& 76.56 & 5.38
& 5.75 & 0.64
& \textbf{5.43} & \textbf{0.58}
& 20.31 & 2.31 \\
& 30\%
& 9.87 & 1.12
& \textbf{7.07} & \textbf{0.79}
& 10.32 & 1.23
& 12.64 & 1.40
& 61.93 & 5.14
& 11.51 & 1.28
& 7.60 & 0.82
& 17.47 & 1.93 \\
& 50\%
& 11.67 & 1.31
& \textbf{9.13} & \textbf{0.90}
& 13.70 & 1.80
& 19.33 & 1.87
& 60.71 & 5.13
& 16.30 & 1.87
& 11.40 & 1.26
& 23.45 & 2.84 \\
\hline

\end{tabular}
\caption{Imputation performance on the curated evaluation benchmark.}
\vspace{-1em}
\label{tab:imputation_results}
\end{table*}

\begin{figure*}[t]
\centering 
\includegraphics[width=0.18\textwidth]{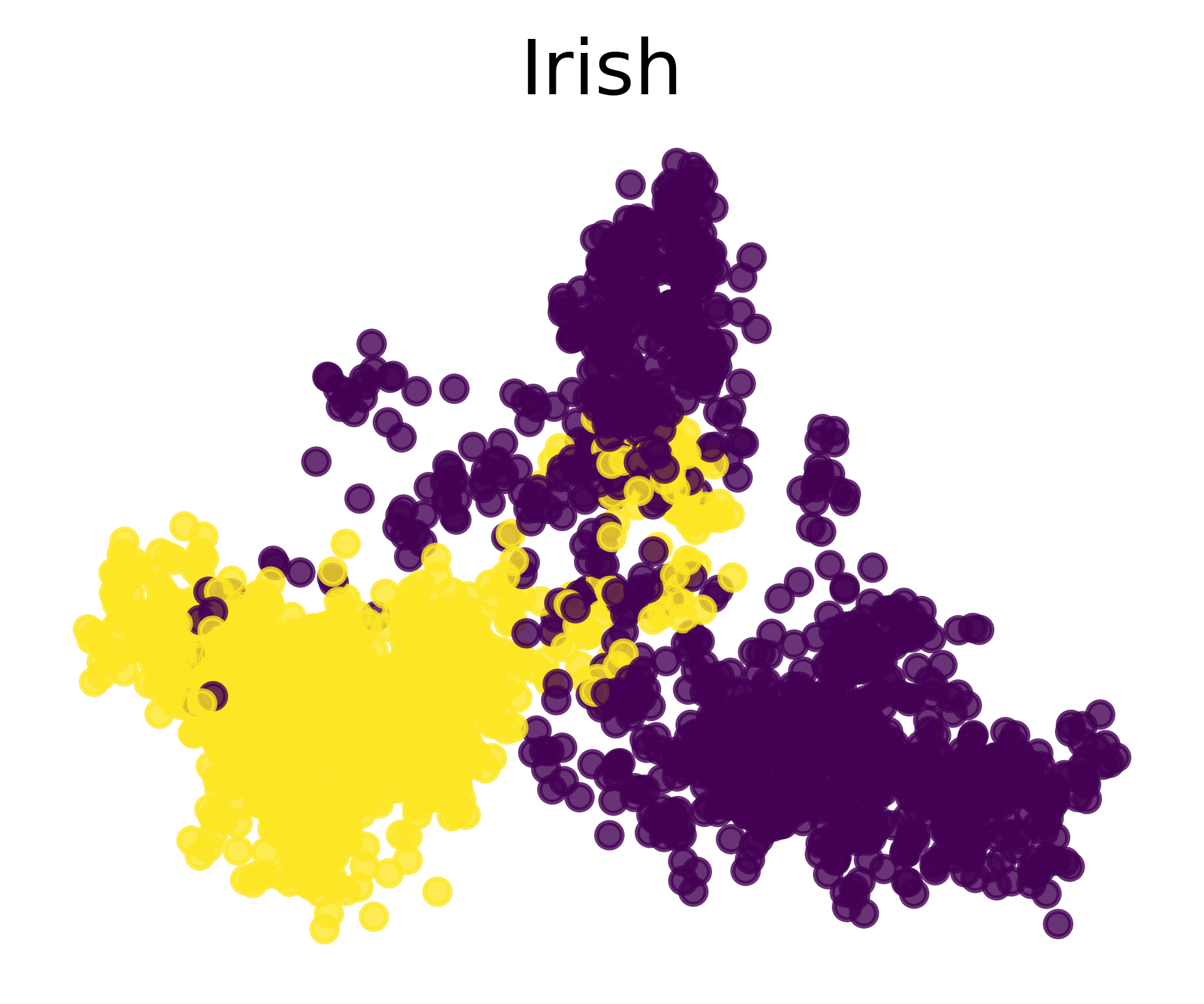} 
\hfill 
\includegraphics[width=0.18\textwidth]{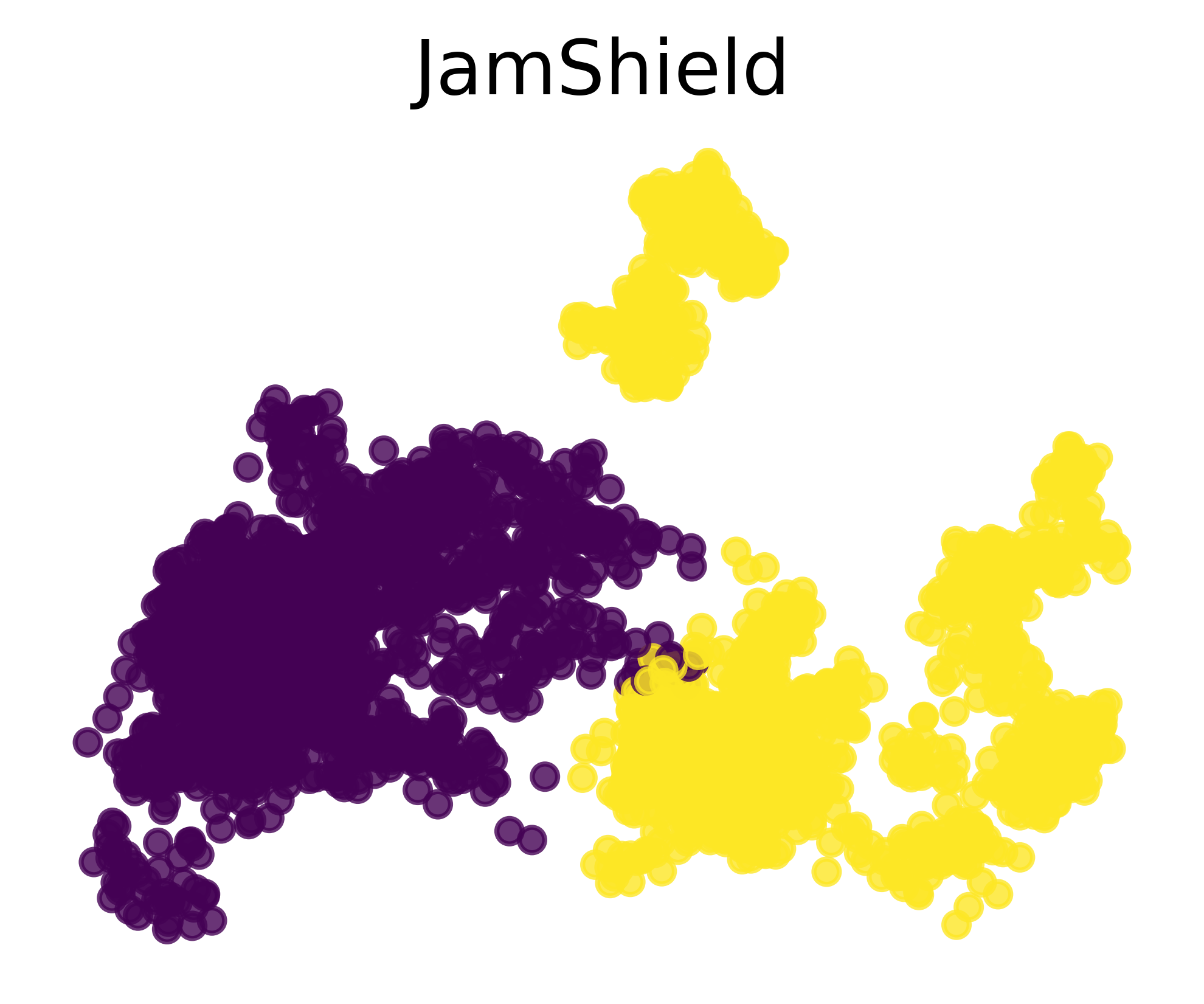} 
\hfill
\includegraphics[width=0.18\textwidth]{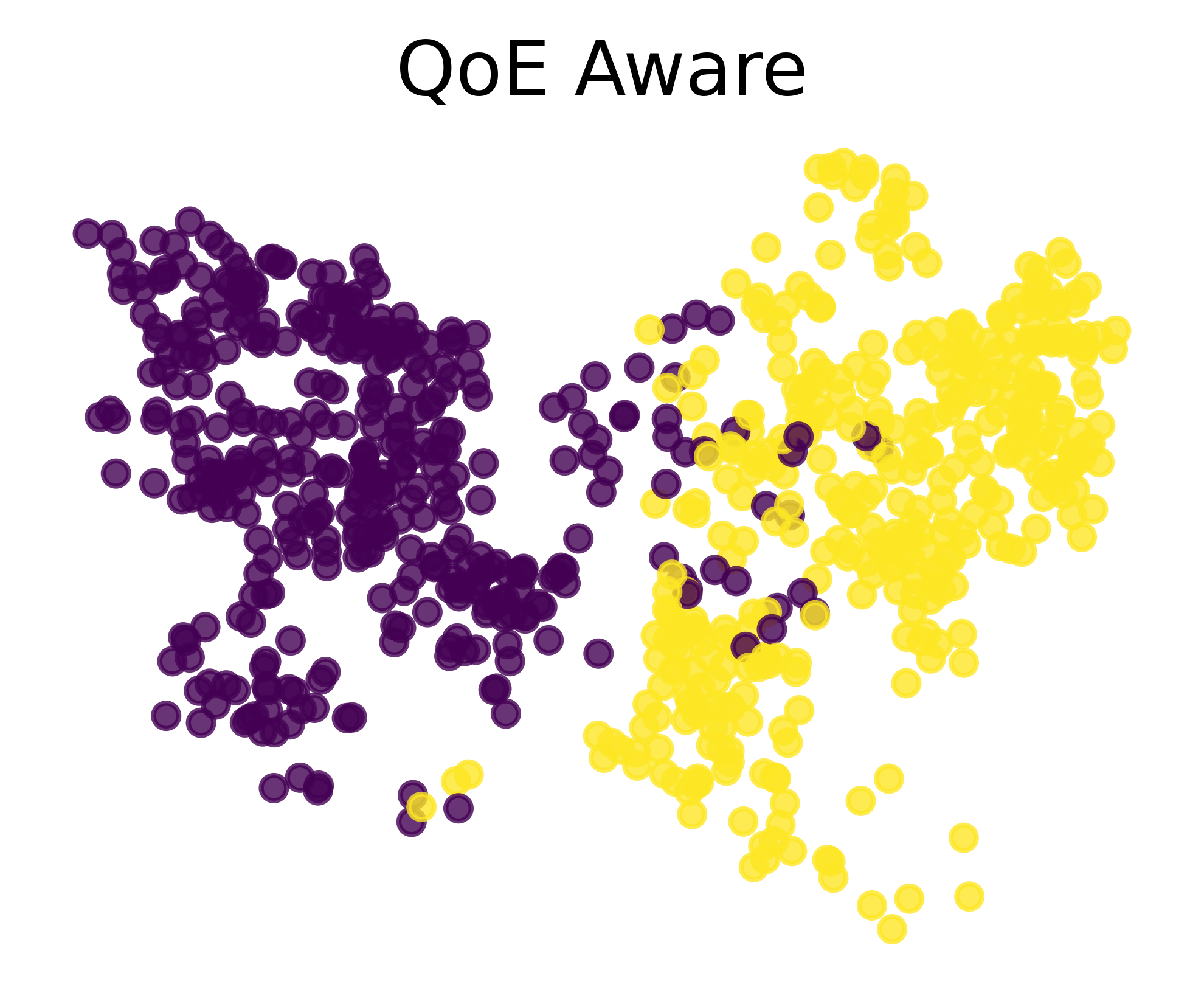} 
\hfill
\includegraphics[width=0.18\textwidth]{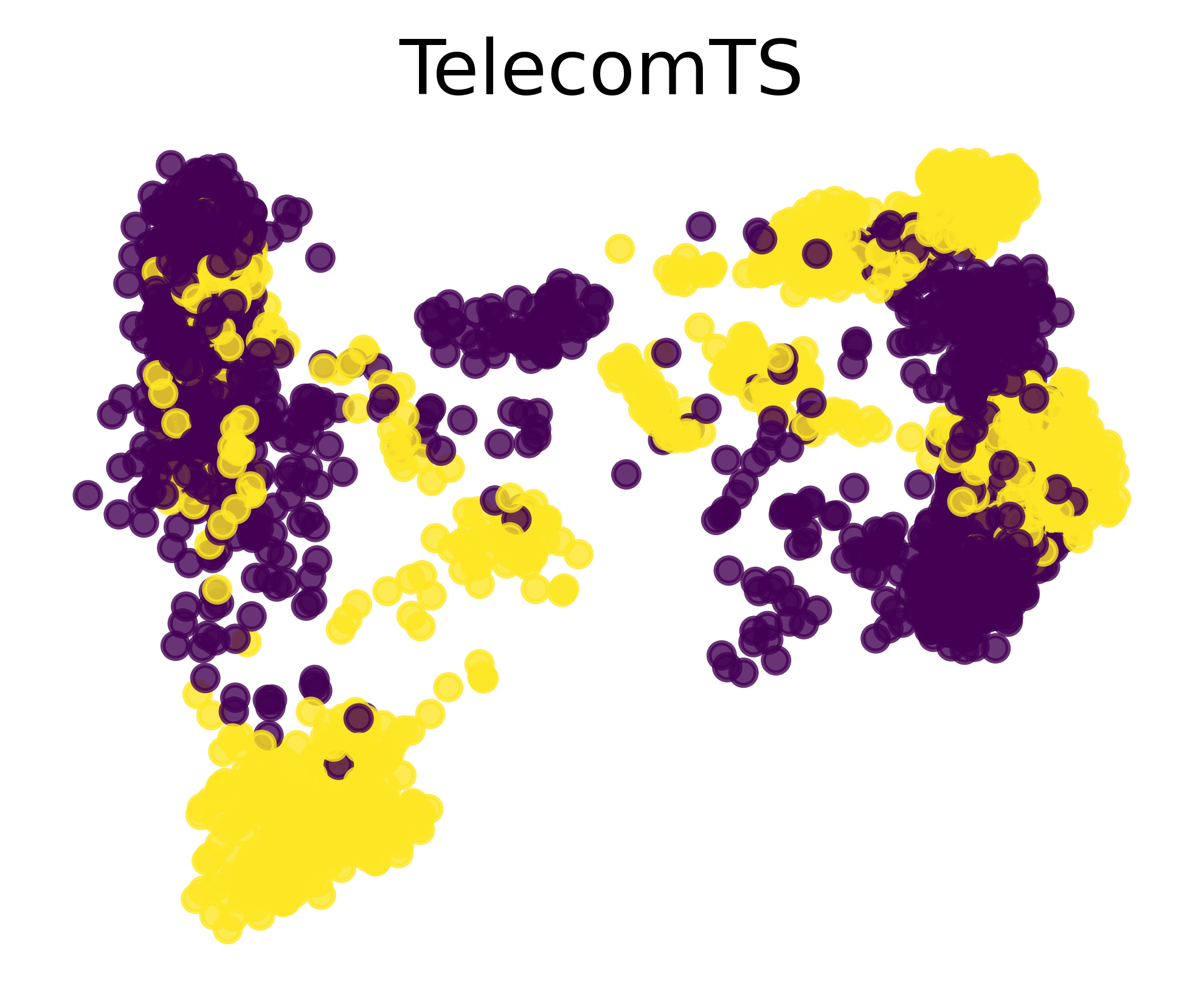}
\hfill
\includegraphics[width=0.18\textwidth]{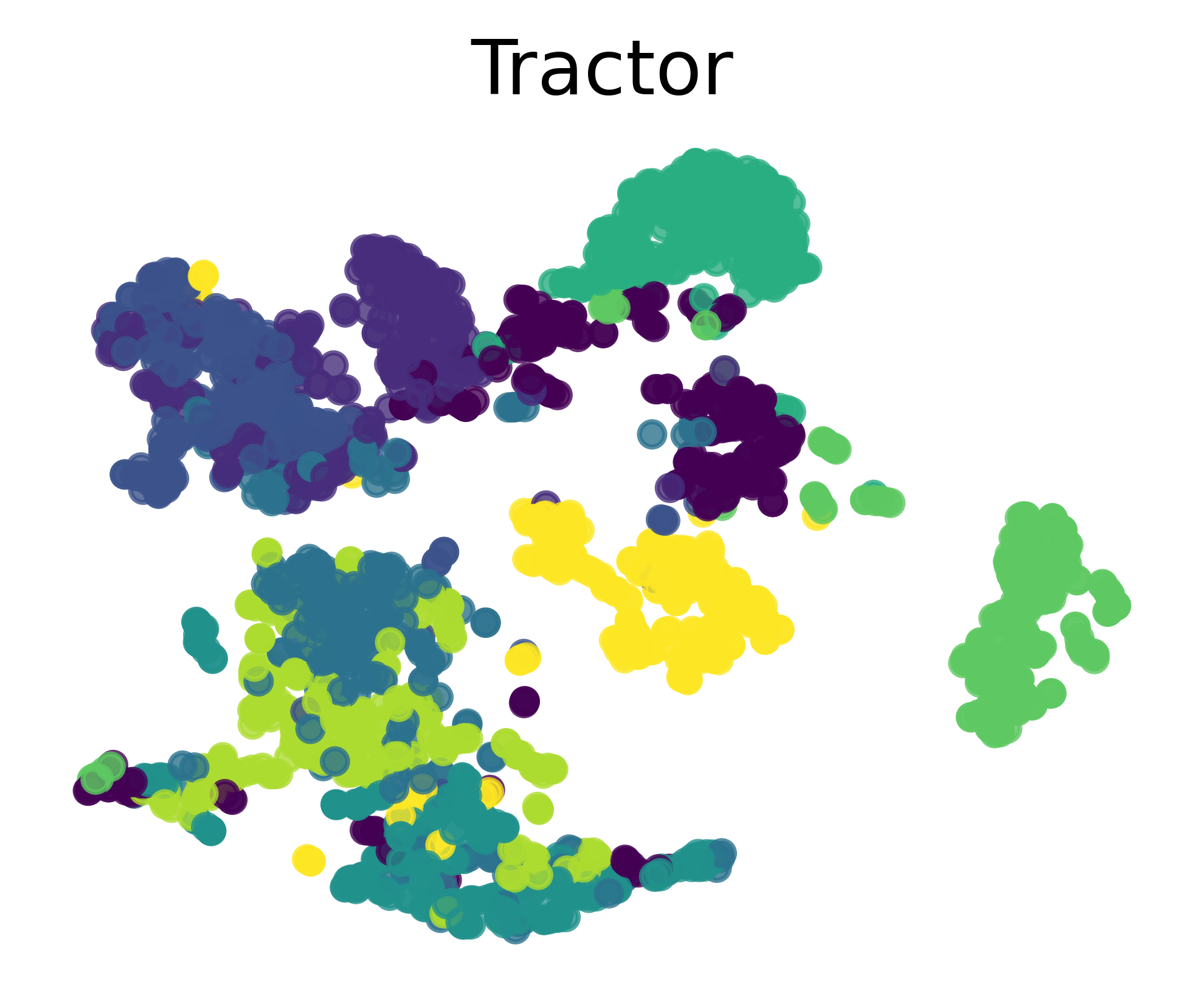} \vspace{-1em} 
\caption{TimeRAN learns separable representations in a zero-shot setting without dataset-specific fine-tuning.} 
\label{fig:representation_learning}
\vspace{-1em}
\end{figure*}

\begin{figure*}[t]
\centering

\begin{subfigure}{0.19\textwidth}
\centering
\includegraphics[width=\linewidth]{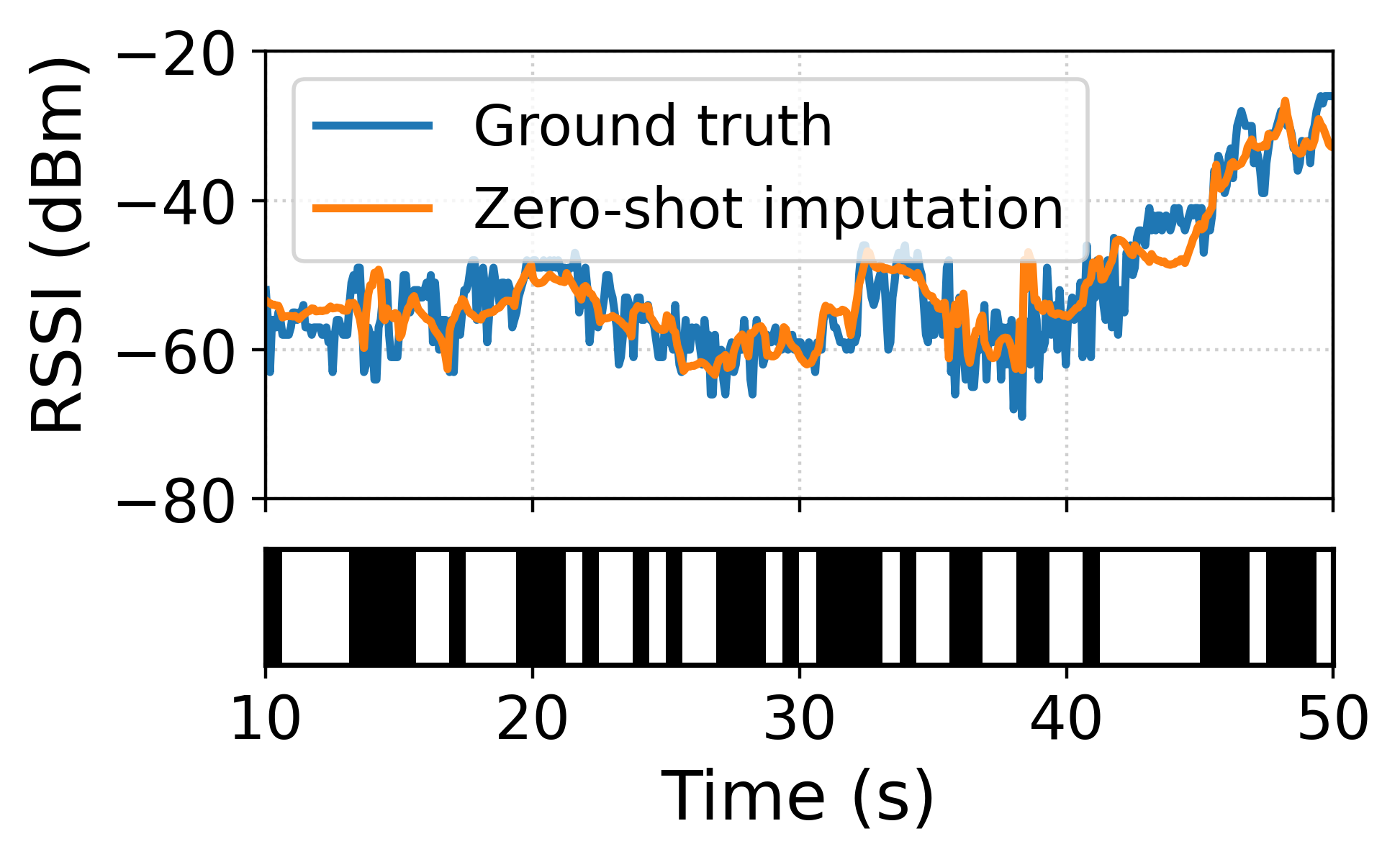}
\vspace{-1.8em}
\caption{WiFi.}
\end{subfigure}
\hfill
\begin{subfigure}{0.19\textwidth}
\centering
\includegraphics[width=\linewidth]{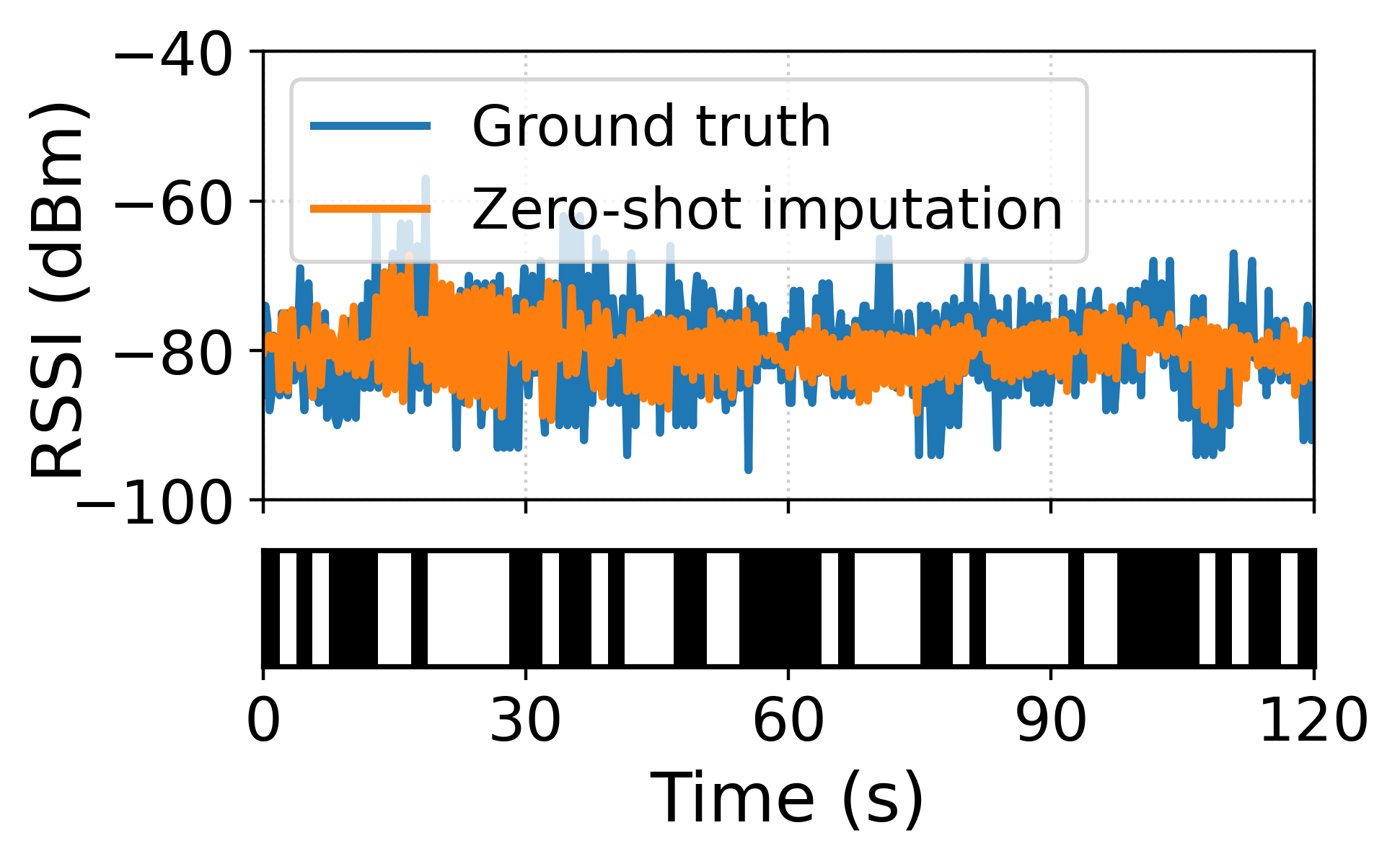}
\vspace{-1.8em}
\caption{Bluetooth.}
\end{subfigure}
\hfill
\begin{subfigure}{0.19\textwidth}
\centering
\includegraphics[width=\linewidth]{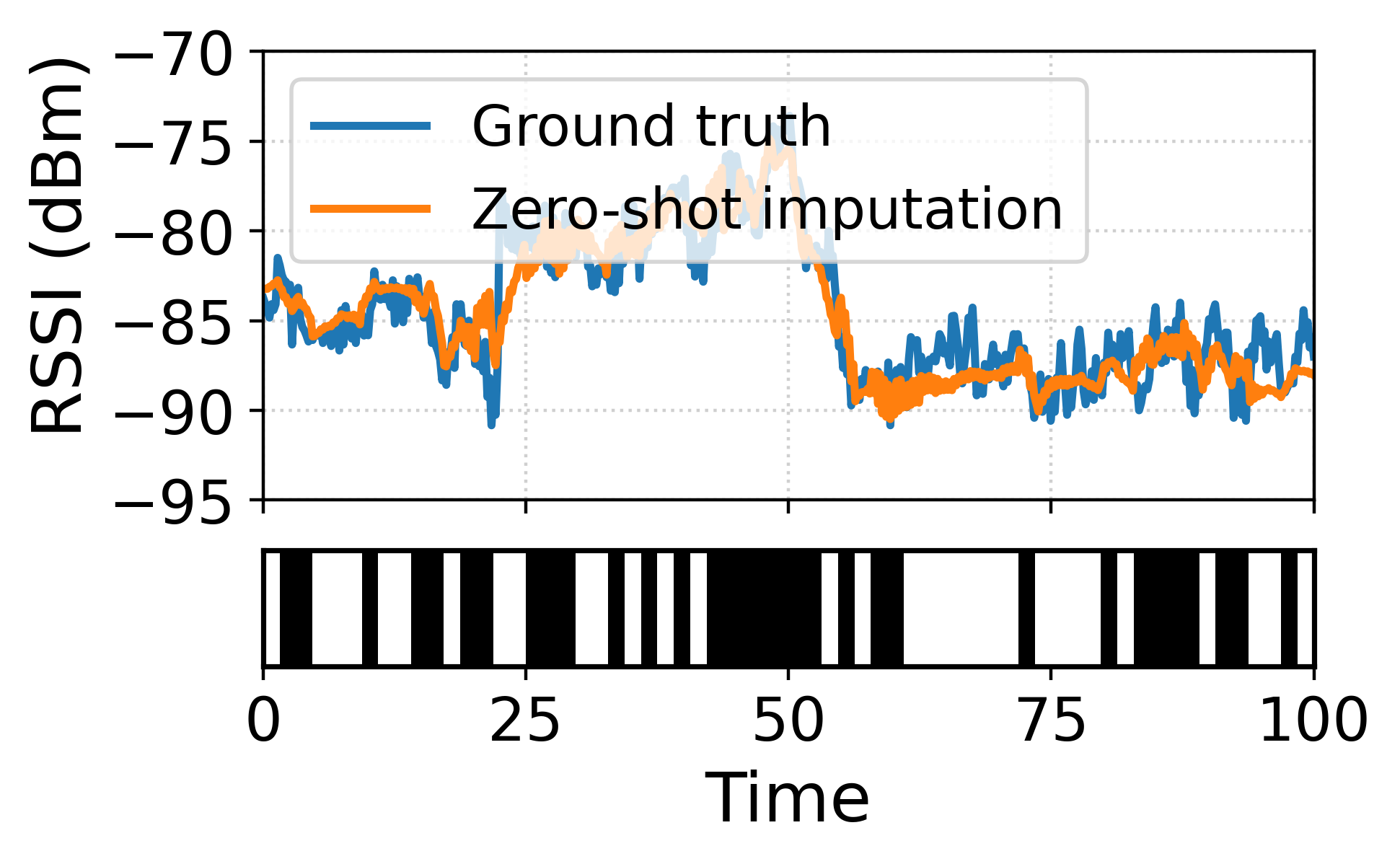}
\vspace{-1.8em}
\caption{LoRa.}
\end{subfigure}
\hfill
\begin{subfigure}{0.19\textwidth}
\centering
\includegraphics[width=\linewidth]{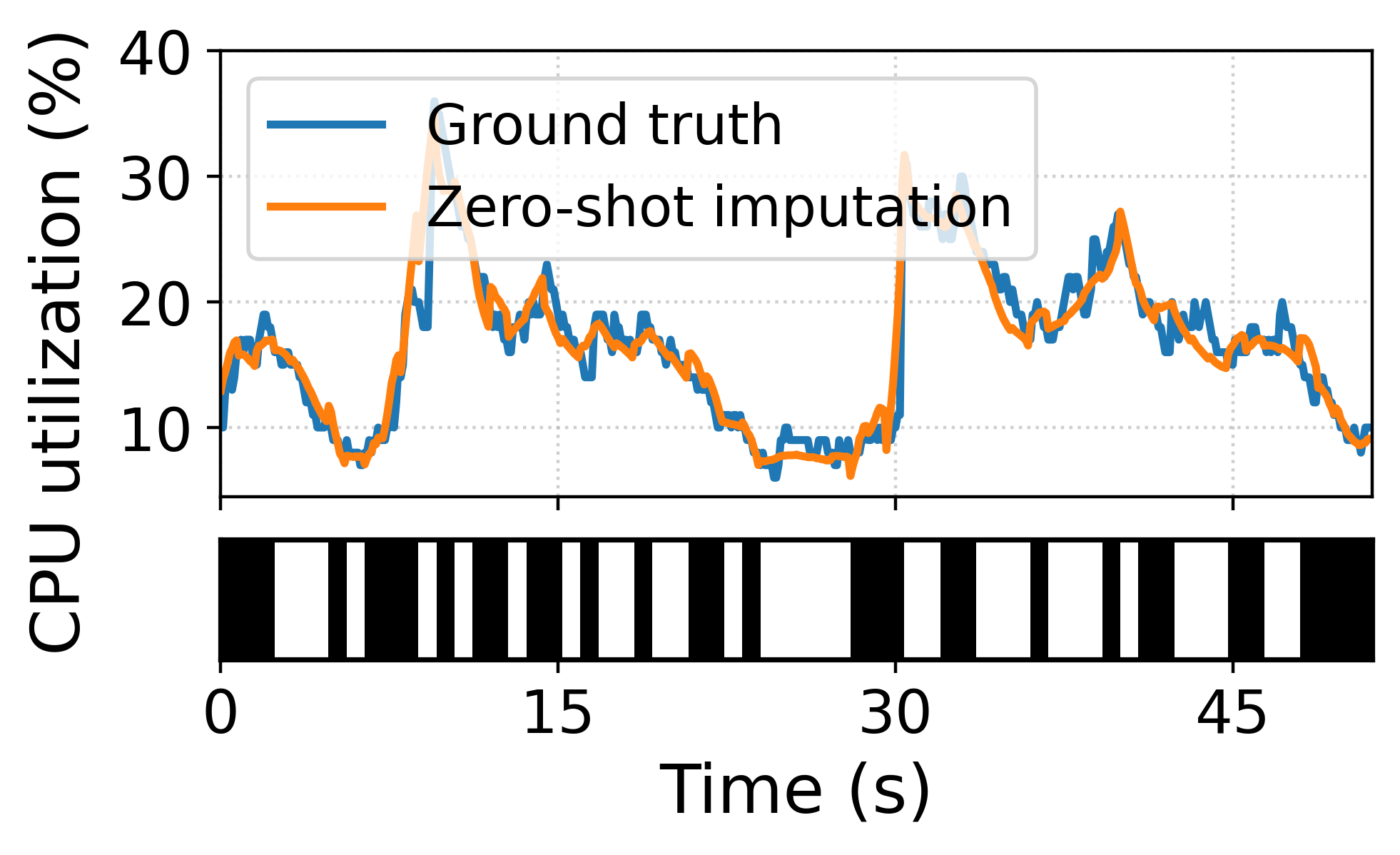}
\vspace{-1.8em}
\caption{CPU utilization.}
\end{subfigure}
\hfill
\begin{subfigure}{0.19\textwidth}
\centering
\includegraphics[width=\linewidth]{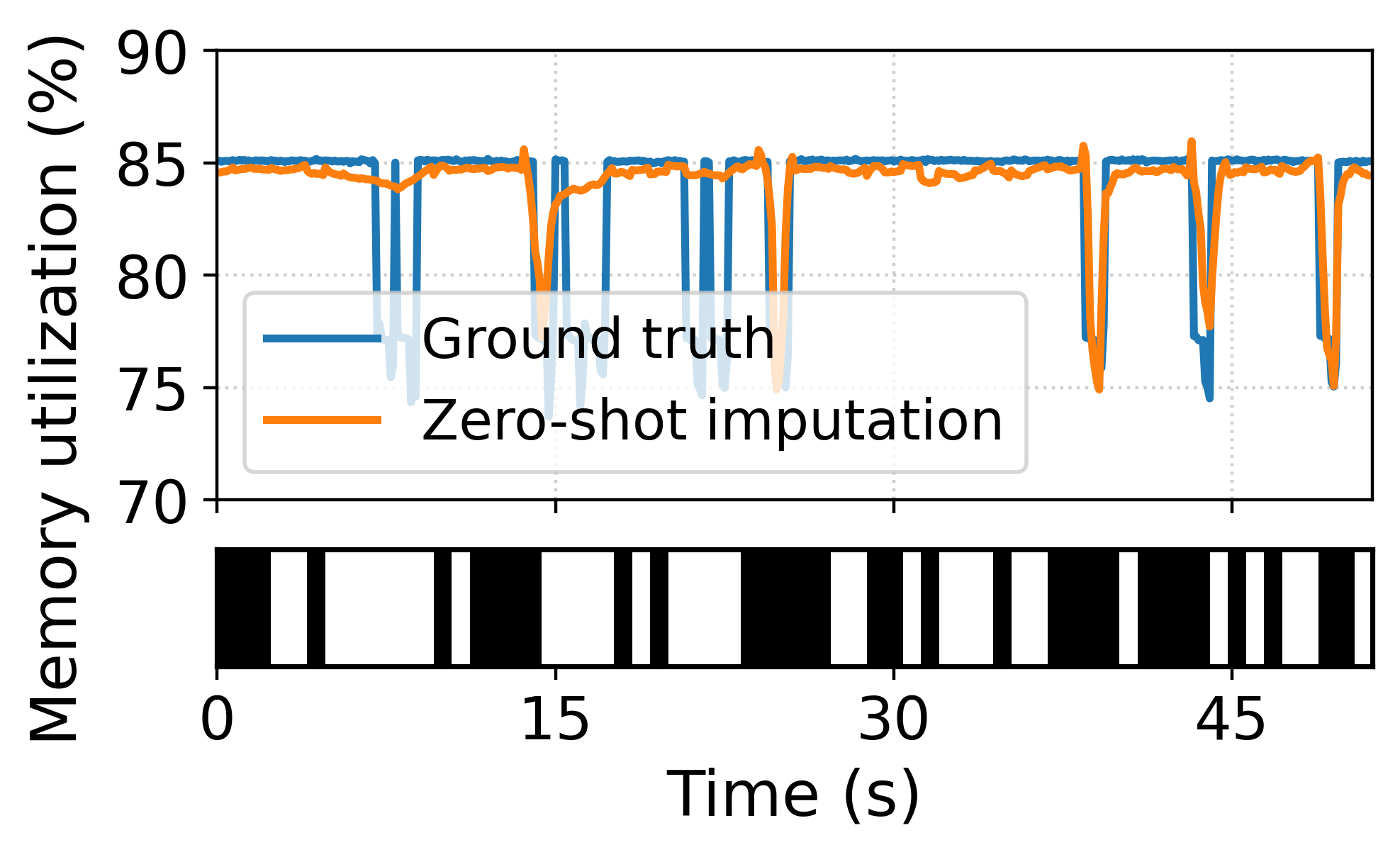}
\vspace{-1.8em}
\caption{Memory utilization.}
\end{subfigure}
\vspace{-1em}
\caption{TimeRAN generalizes to unseen environments in a zero-shot setting.}
\label{fig:zero_shot_imputation_all}
\vspace{-0.8em}
\end{figure*}

\begin{figure*}[t]
\centering

\begin{subfigure}{0.47\textwidth}
\centering
\includegraphics[width=\linewidth]{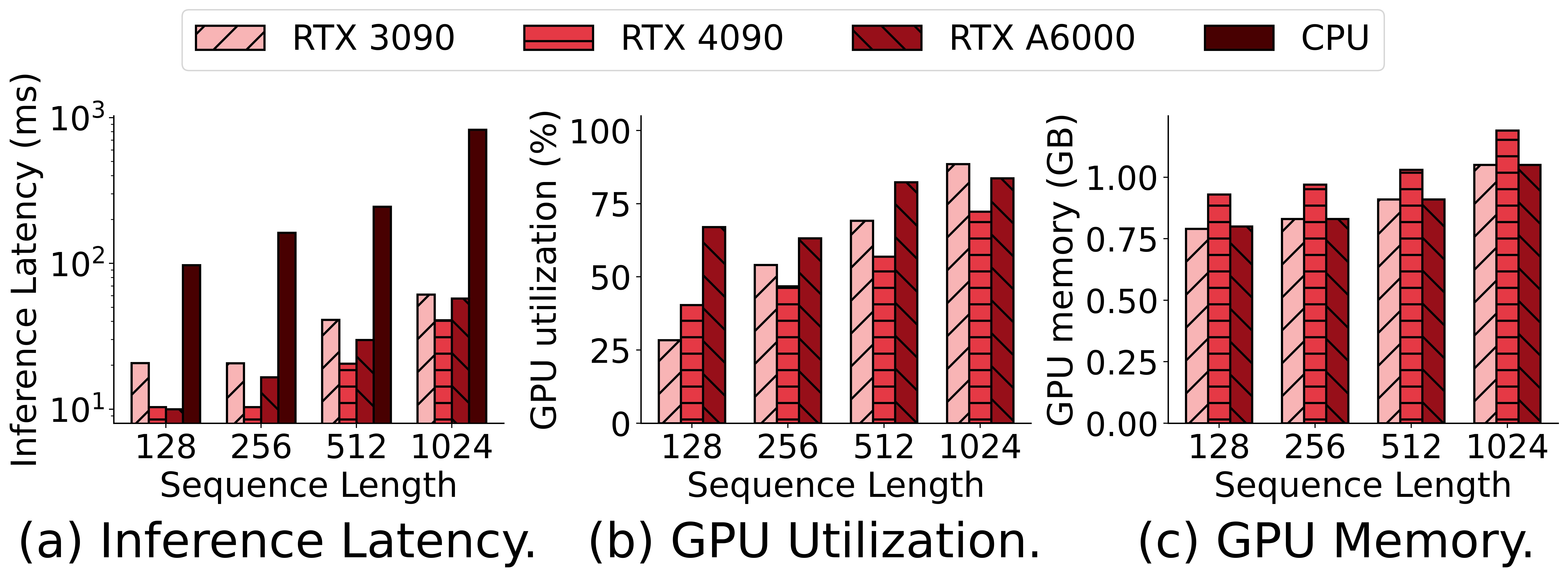}
\vspace{-1.6em}
\caption{Impact of input window size.}
\end{subfigure}
\hfill
\begin{subfigure}{0.47\textwidth}
\centering
\includegraphics[width=\linewidth]{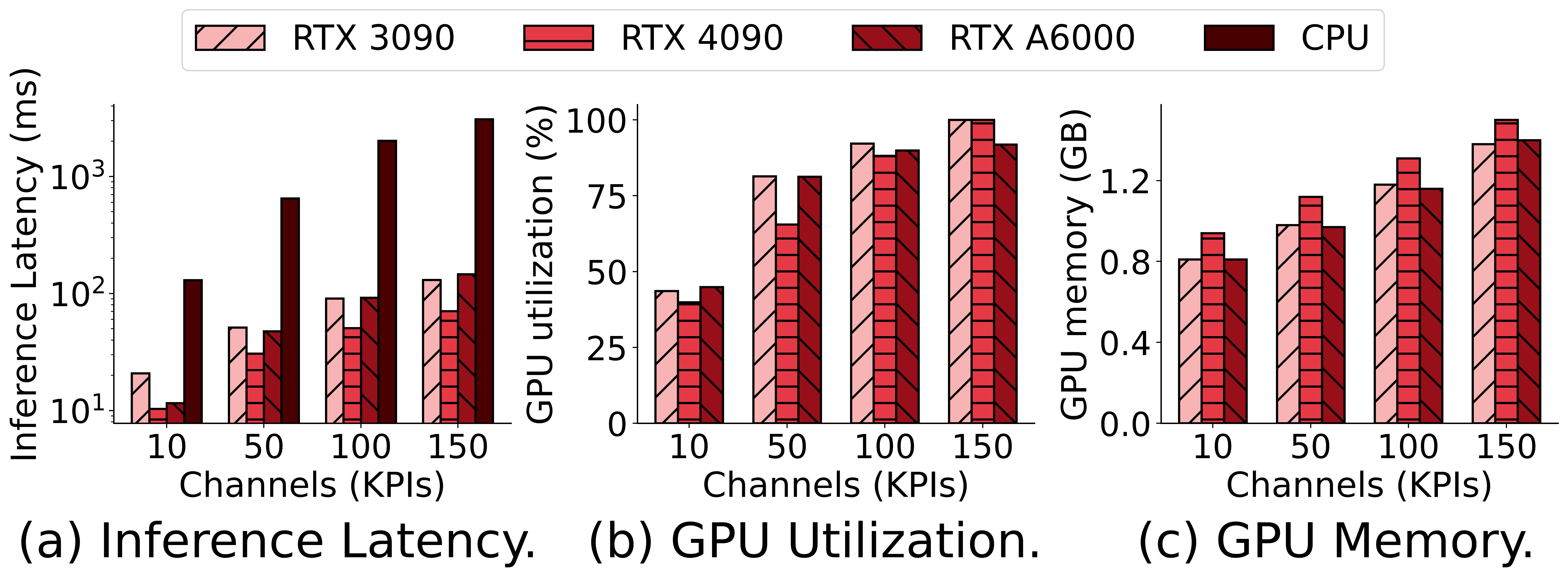}
\vspace{-1.6em}
\caption{Impact of channel dimensionality.}
\end{subfigure}
\vspace{-1em}
\caption{System-level performance of TimeRAN under varying input characteristics during inference.}
\label{fig:system_results}
\vspace{-1em}
\end{figure*}

\subsection{Generalization Analysis}
\label{sec:generalization_analysis}

\noindent We next evaluate the ability of \texttt{TimeRAN} to learn representations that generalize across unseen environments. 
In particular, we examine both (i) the separability of learned representations without task-specific fine-tuning and (ii) the effectiveness of zero-shot downstream task inference in domains that differ from those observed during training.

\noindent{\textbf{8.2.1\quad Representation Separability.}} Fig.~\ref{fig:representation_learning} visualizes the latent representations learned by \texttt{TimeRAN} across several datasets using dimensionality reduction techniques, specifically PCA \cite{pca}, to project the learned high-dimensional embeddings into a lower-dimensional space. Despite operating in a zero-shot setting, the learned embeddings exhibit clear clustering structures corresponding to different classes. For example, mobility states, anomalies, and service types form well-separated clusters across the Irish \cite{irish_dataset}, JamShield \cite{jamshield_dataset}, QoE Aware \cite{qoe_aware_dataset}, TelecomTS \cite{telecomts_dataset}, and Tractor \cite{tractor_dataset} datasets. These results indicate that \texttt{TimeRAN} learns transferable temporal representations that enable simple linear classifiers to separate distinct data patterns in the latent space.

\noindent{\textbf{8.2.2\quad Zero-Shot Generalization to Unseen Environments.}} We further assess the zero-shot generalization capability of \texttt{TimeRAN} on downstream tasks using data not encountered during either pre-training or fine-tuning, focusing on imputation as a representative case. Fig.~\ref{fig:zero_shot_imputation_all} presents results across diverse datasets spanning multiple wireless environments, including WiFi \cite{wifi}, Bluetooth \cite{ble}, and LoRa \cite{lora_dataset}, as well as system-level telemetry collected from the base station (Distributed Unit) of our testbed. These datasets are selected to cover domains with temporal characteristics similar to RAN telemetry as well as fundamentally different sources, as described in ~\cite{wifi, ble, lora_dataset}. In each case, we randomly mask $50\%$ of the input time series and reconstruct the missing segments (white regions). Across all scenarios, \texttt{TimeRAN} accurately recovers the masked portions, closely matching the original input time series.

\noindent Overall, these results demonstrate that \texttt{TimeRAN} learns transferable representations that  can generalize across heterogeneous environments and extend beyond the RAN domain.

\subsection{System Overhead During Inference}
\label{sec:system_overhead}

\noindent In addition, we evaluate the resource overhead introduced by \texttt{TimeRAN} by examining how the input window size $T$ and channel dimensionality $C$ impact inference latency and overall resource utilization across different computing platforms.

\noindent{\textbf{8.3.1\quad Impact of Input Window Size.}}
Fig.~\ref{fig:system_results}(a) depicts system performance as the input window size $T$ increases from $128$ to $1024$, with the number of input channels fixed at $C=30$. As expected, inference latency increases with window size as the self-attention layers must process longer temporal contexts. For example, latency on the RTX A6000 increases from approximately $30$ ms at window size $128$ to about $70$ ms at $1024$, while CPU inference exceeds $800$ ms. GPU utilization rises from roughly $30\%$ to over $90\%$, reflecting the increased compute workload. GPU memory consumption also increases gradually with window size but remains relatively modest, reaching approximately $1.1$ GB.  CPU memory remains largely unchanged, with ~1 core used for data transfer, as computation is offloaded to the GPU.

\noindent{\textbf{8.3.2\quad Impact of Input Channel Dimensionality.}}
Fig.~\ref{fig:system_results}(b) illustrates the effect of increasing the number of input channels $C \in \{10,50,100,150\}$ while keeping the window size fixed ($T=512$). As channel dimensionality increases, inference latency and GPU utilization rise due to the larger input representation processed by \texttt{TimeRAN}. For instance, latency on the RTX A6000 increases from roughly $40$\,ms at $10$ channels to around $120$\,ms at $150$ channels, while CPU latency grows to over $2000$\,ms. GPU utilization approaches $100\%$ at higher dimensionalities, and GPU memory usage increases to approximately $1.3$,GB, remaining relatively modest overall.

\section{EVALUATION IN REAL 5G TESTBED}
\label{sec:integration}

\noindent Finally,  we integrate \texttt{TimeRAN} into our proof-of-concept 5G testbed and evaluate it across all supported RAN downstream tasks, selecting a representative real-world use case scenario for each task. All subsequent experiments are conducted using live telemetry collected directly from our private 5G network deployment, where over 20 cross-layer RAN Key Performance Indicators (KPIs) are continuously monitored at a fine-grained temporal granularity of \(10\,\mathrm{ms}\). Across all these tasks, \texttt{TimeRAN} processes the telemetry using a sliding window of length \(T=512\) (corresponding to \(5.12\,\mathrm{s}\)).

\begin{figure}[t]
\centering
\begin{subfigure}[t]{0.46\columnwidth}
    \includegraphics[width=\linewidth]{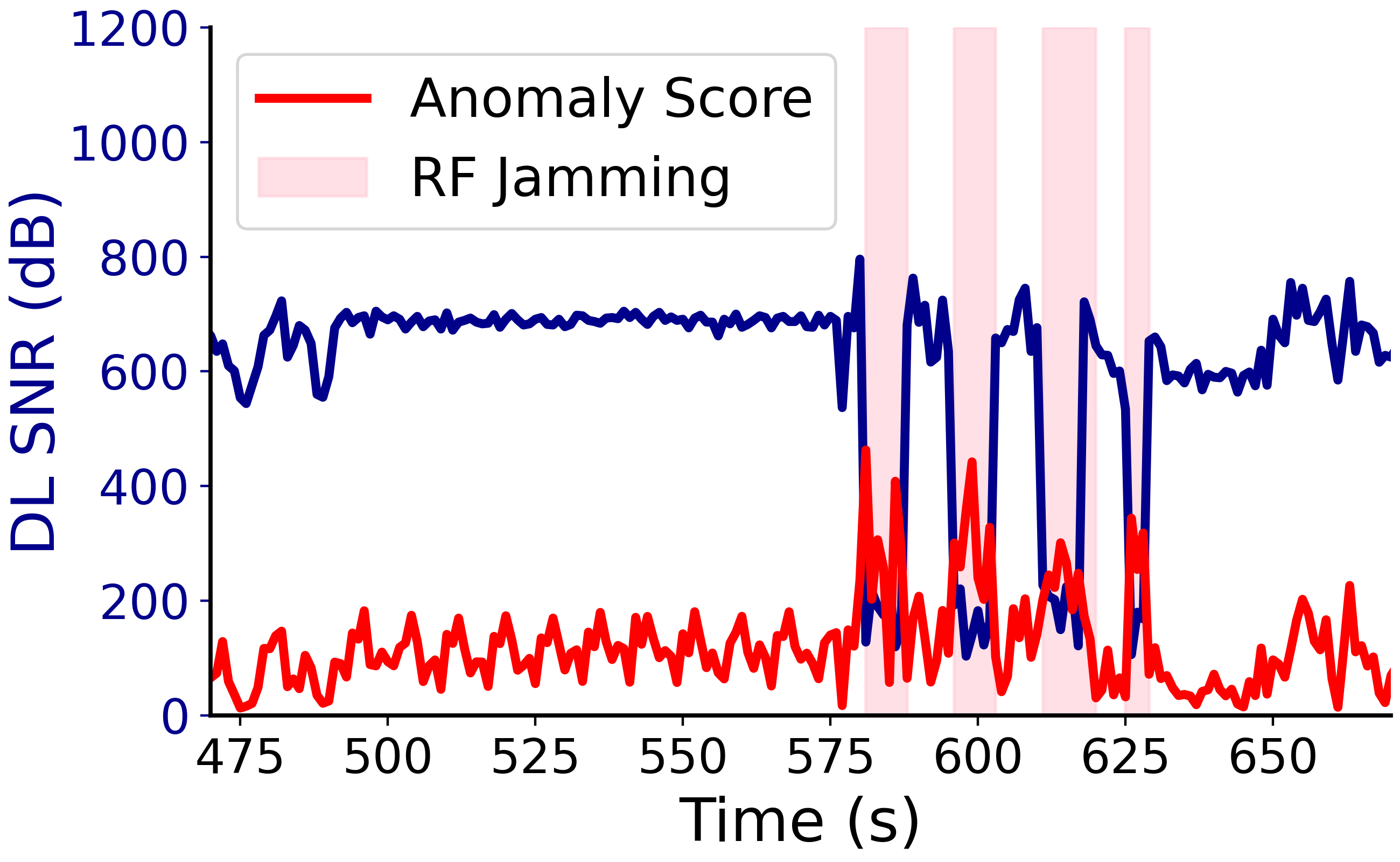}
    \vspace{-1.8em}
    \caption{Anomaly detection.}
\end{subfigure}
\begin{subfigure}[t]{0.46\columnwidth}
    \includegraphics[width=\linewidth]{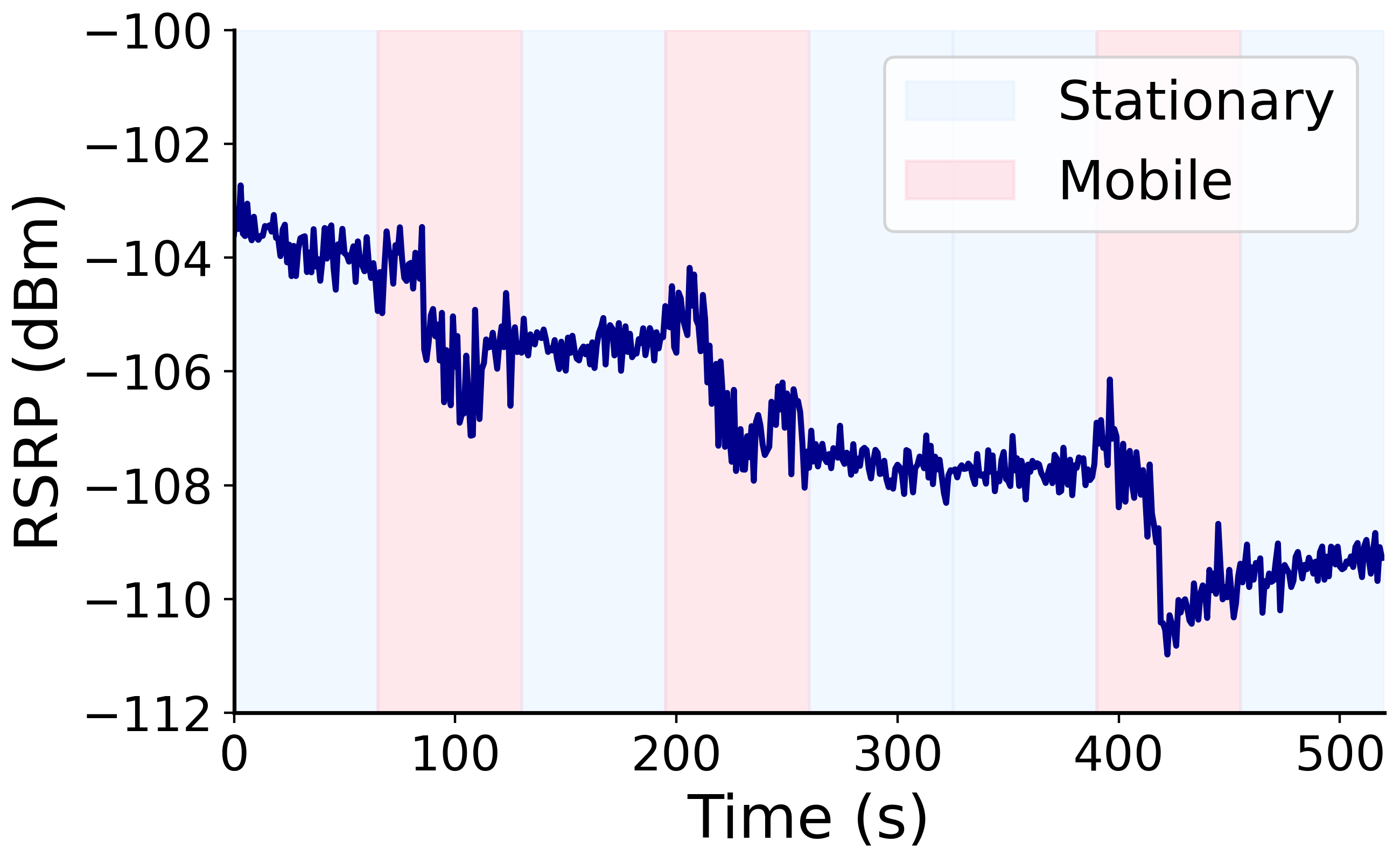}
    \vspace{-1.8em}
    \caption{Classification.}
\end{subfigure}
\begin{subfigure}[t]{0.46\columnwidth}
    \includegraphics[width=\linewidth]{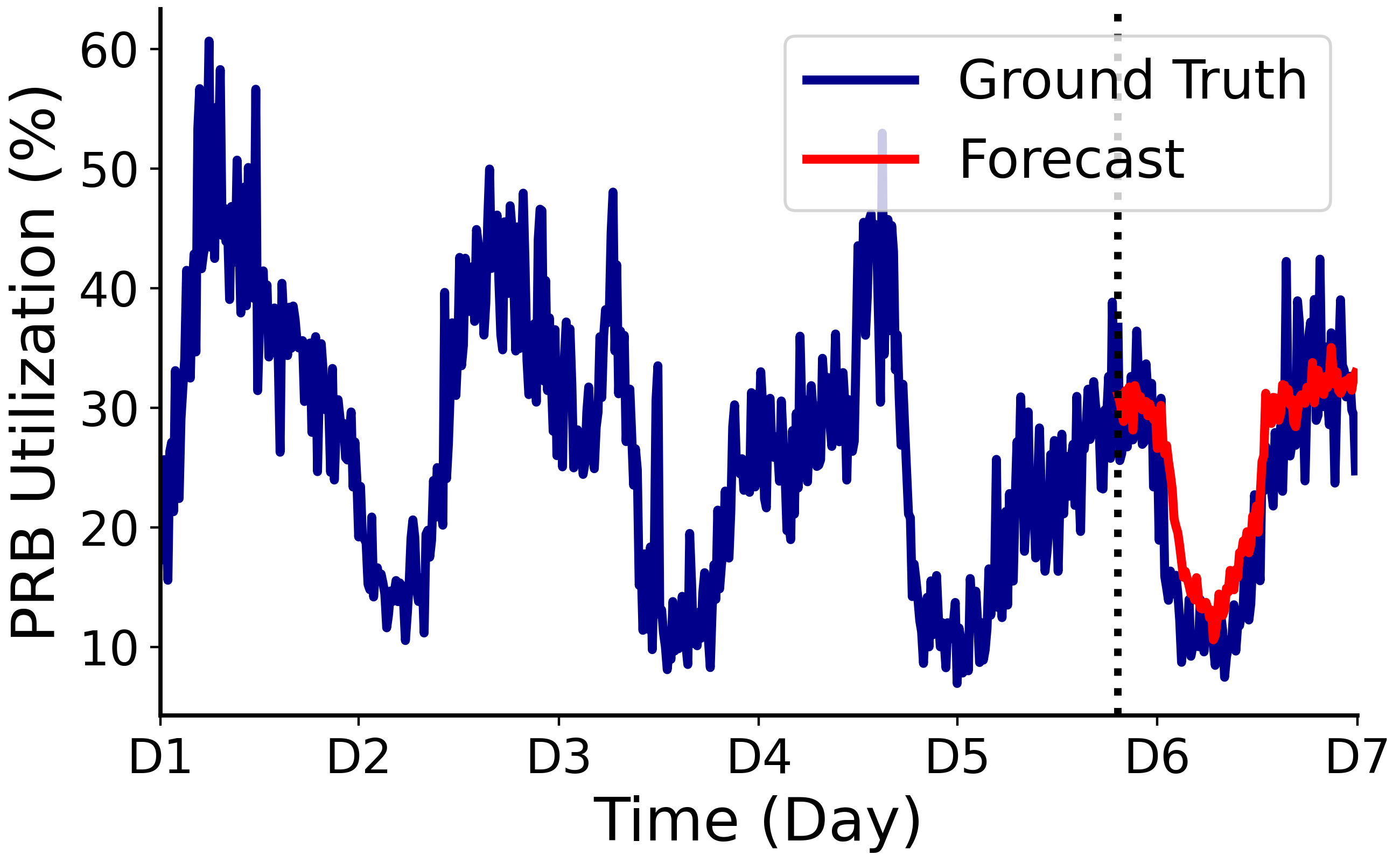}
    \vspace{-1.8em}
    \caption{Forecasting.}
\end{subfigure}
\begin{subfigure}[t]{0.46\columnwidth}
    \includegraphics[width=\linewidth]{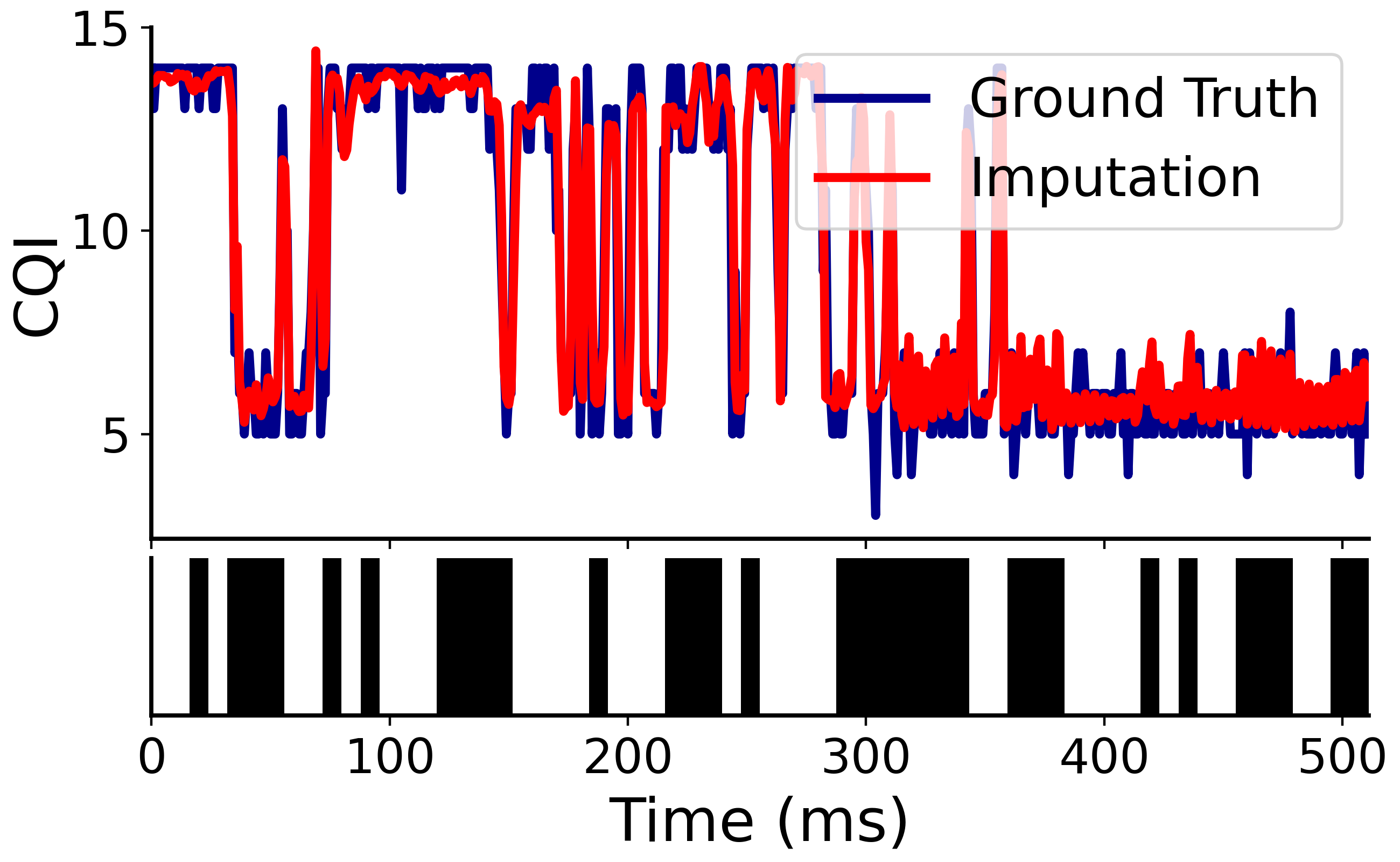}
    \vspace{-1.8em}
    \caption{Imputation.}
\end{subfigure}
\vspace{-1em}
\caption{TimeRAN performance across representative RAN analytics tasks on our over-the-air 5G testbed.}
\label{fig:testbed_results}
\vspace{-1.8em}
\end{figure}

\noindent{\textbf{9.1.1\quad Anomaly Detection.}} For anomaly detection, we induce a realistic RF jamming attack using a USRP X310 SDR positioned approximately \(5\,\mathrm{m}\) from the mobile devices, transmitting over-the-air interference on the data channels at maximum gain. \texttt{TimeRAN} produces a normalized anomaly score in the range $[0,1]$, where higher values correspond to abnormal network behavior. As shown in Fig.~\ref{fig:testbed_results}(a), the anomaly score sharply increases during the jamming attack, accurately identifying it in real time.

\noindent{\textbf{9.1.2 \quad  Classification.}} For classification, we consider a mobility detection task in which a single UE moves within the laboratory (approximately $80\,\mathrm{m}^2$), with the goal of classifying the user state as either stationary or mobile. By jointly analyzing RSRP measurements and additional PHY-layer telemetry indicators, \texttt{TimeRAN} reliably distinguishes between stationary and mobile user states, as illustrated in Fig.~\ref{fig:testbed_results}(b).

\noindent{\textbf{9.1.3 \quad Forecasting.}} For forecasting, we consider a scenario in which all mobile devices are attached to an enhanced Mobile Broadband (eMBB) slice while generating downlink traffic through YouTube video streaming. The objective is to predict the future network load at the base station, measured in terms of Physical Resource Block (PRB) utilization. Using historical telemetry observations, \texttt{TimeRAN} accurately forecasts future PRB utilization and reliably captures temporal fluctuations in network load, as illustrated in Fig.~\ref{fig:testbed_results}(c).

\noindent{\textbf{9.1.4\quad Imputation.}} For imputation, we consider a scenario in which a user with a mobile device moves within the laboratory while Channel Quality Indicator (CQI) measurements are partially masked from the telemetry stream with a masking ratio of 50\% (white regions). \texttt{TimeRAN} reconstructs the missing values from the partially observed time series. As shown in Fig.~\ref{fig:testbed_results}(d), the model accurately recovers the masked segments while preserving the temporal CQI dynamics.

 These results demonstrate that \texttt{TimeRAN} reliably supports diverse RAN tasks directly from live telemetry, enabling real-time inference within operational 5G deployments.

\section{CONCLUSION}
\noindent In this work, we introduced \texttt{TimeRAN}, a unified multi-task learning framework for the convergence of RAN downstream tasks. Leveraging large-scale pretraining on the \textit{TimeRAN DataPile}, \texttt{TimeRAN} learns transferable representations that generalize across diverse tasks and environments with minimal supervision. Extensive evaluations and real-world testbed results demonstrate state-of-the-art performance alongside low-latency inference with minimal resource overhead.

\renewcommand{\refname}{REFERENCES}
\bibliographystyle{ACM-Reference-Format}
\bibliography{sample-base}

\end{document}